%% file: KinematicQuenching.tex
\documentclass[useAMS,usenatbib,10pt]{mnras}
\usepackage{listings}
\usepackage{times}  
\usepackage{graphics}
\usepackage{subfigure}
\usepackage{amssymb}
\usepackage{hyperref}
\usepackage{aas_macros}
\usepackage{lscape}
\usepackage{rotating}
\usepackage[export]{adjustbox}
\usepackage{graphicx}
\def\gsim{\ga}
\def\lsim{\la}

\def\eagle{{\sc eagle}}
\def\Hydrangea{{\sc hydrangea}}
\def\gsim{ \lower .75ex \hbox{$\sim$} \llap{\raise .27ex \hbox{$>$}} }
\def\lsim{ \lower .75ex \hbox{$\sim$} \llap{\raise .27ex \hbox{$<$}} }
\def\simprop{ \lower .75ex \hbox{$\sim$} \llap{\raise .27ex \hbox{$\propto$}} }
\title[Connecting kinematics, mass and environment]{
The connection between mass, environment and slow rotation in simulated galaxies}
\author[Claudia del P. Lagos et al.]{
\parbox[t]{\textwidth}{
\vspace{-1.0cm}
Claudia del P. Lagos$^{1,2}$\thanks{E-mail: claudia.lagos@icrar.org}, Joop Schaye$^{3}$, Yannick Bah\'e$^{3}$, Jesse Van de Sande$^{4}$, Scott T. Kay$^{5}$, David Barnes$^{5,6}$, Timothy A. Davis$^{7}$, Claudio Dalla Vecchia$^{8,9}$}
\vspace*{6pt} \\
$^{1}$International Centre for Radio Astronomy Research (ICRAR), M468, University of Western Australia, 35 Stirling Hwy, Crawley, WA 6009, Australia.\\
$^{2}$Australian Research Council Centre of Excellence for All-sky Astrophysics (CAASTRO), 44 Rosehill Street Redfern, NSW 2016, Australia.\\
$^{3}$Leiden Observatory, Leiden University, PO Box 9513, NL-2300 RA Leiden, the Netherlands.\\
$^{4}$Sydney Institute for Astronomy, School of Physics A28, The University of Sydney, NSW 2006, Australia.\\
$^{5}$Jodrell Bank Centre for Astrophysics, School of Physics and Astronomy, The University of Manchester, Manchester M13 9PL, UK.\\
$^{6}$Department of Physics, Kavli Institute for Astrophysics and Space Research, Massachusetts Institute of Technology, Cambridge, MA 02139, USA.\\
$^{7}$Department of Physics and Astronomy, Cardiff University, Queens Buildings, The Parade, Cardiff CF24 3AA, United Kingdom.\\
$^{8}$Instituto de Astrof\'isica de Canarias, C/V\'ia L\'actea s/n, E-38205 La Laguna, Tenerife, Spain.\\
$^{9}$Departamento de Astrof\'isica, Universidad de La Laguna, Av. del Astrof\'isico Francisco S\'anchez s/n, E-38206 La Laguna, Tenerife, Spain.
\vspace*{-0.5cm}}

\begin{document}


\pagerange{\pageref{firstpage}--\pageref{lastpage}} \pubyear{2017}

\maketitle

\label{firstpage}

\begin{abstract}
Recent observations from integral field spectroscopy (IFS) indicate that the fraction of galaxies that are
slow rotators, $F_{\rm SR}$, depends primarily on stellar mass, with no significant dependence on environment. We investigate these trends and the formation paths of slow rotators (SRs) using the \eagle\ and \Hydrangea\ hydro-dynamical simulations. \eagle\ consists of several cosmological boxes of volumes up to $(100\,\rm Mpc)^3$, while \Hydrangea\ consists of $24$ cosmological simulations of galaxy clusters and their environment. Together they provide a statistically significant sample in the stellar mass range $10^{9.5}\,\rm M_{\odot}-10^{12.3}\,\rm M_{\odot}$, of $16,358$ galaxies. We construct IFS-like cubes and measure stellar spin parameters, $\lambda_{\rm R}$, and ellipticities, allowing us to classify galaxies into slow/fast rotators as in observations. The simulations display a primary dependence of $F_{\rm SR}$ on stellar mass, with a weak dependence on environment. At fixed stellar mass, satellite galaxies are more likely to be SRs than centrals. $F_{\rm SR}$ shows a dependence on halo mass at fixed stellar mass for central galaxies, while no such trend is seen for satellites. We find that $\approx 70$\% of SRs at $z=0$ have experienced at least one merger with mass ratio $\ge 0.1$, with dry mergers being at least twice more common than wet mergers. Individual dry mergers tend to decrease $\lambda_{\rm R}$, while wet mergers mostly increase it. However, $30$\% of SRs at $z=0$ have not experienced mergers, and those inhabit halos with median spins twice smaller than the halos hosting the rest of the SRs. Thus, although the formation paths of SRs can be varied, dry mergers and/or halos with small spins dominate.
\end{abstract}


\section{Introduction}

Integral field spectroscopy (IFS) is opening a new window
for exploring galaxy formation and evolution.
Many recent surveys, such as ATLAS$^{\rm 3D}$ \citep{Cappellari11},
SAMI \citep{Croom12,Bryant15}, CALIFA \citep{Sanchez12},
MASSIVE \citep{Ma14} and MaNGA \citep{Bundy15}
are exploring how the resolved kinematics of the stars and ionised gas
relate to global galaxy properties, such as stellar mass,
colour, star formation rate (SFR) and environment, among others.
The most revolutionizing aspect of these surveys is that due to their
significant volumes, they are
able to observe many hundreds to many thousands galaxies spanning
a very wide dynamic range in mass and environment. This enables
the galaxy population to be dissected into many properties, but most
significantly into stellar and environment, which are thought to be primary
drivers in the evolution of galaxies (e.g. \citealt{Peng10}).

One of the most prominent early examples of the success of IFS surveys was
the pioneering work of the SAURON \citep{Bacon01} and ATLAS$^{\rm 3D}$ \citep{Cappellari11} surveys,
comprised of $260$ early-type galaxies in total.
These surveys showed that the stellar kinematics and distributions of
stars are not strongly correlated in early-types, and thus that morphology
is not necessarily a good indicator of the dynamics of galaxies \citep{Krajnovic13}.
Based on these surveys, \citet{Emsellem07,Emsellem11} coined the
terms {\it slow} and {\it fast} rotators, and proposed
the $\lambda_{\rm R}$ parameter, which measures how rotationally or dispersion-dominated
a galaxy is,
as a new, improved scheme to classify galaxies.
The most significant trend found by \citet{Emsellem11} and extended recently to higher stellar masses
by \citet{Veale17}, is that the fraction of slow rotators increases steeply with stellar mass and
towards denser environments, and that
the vast majority of S0 galaxies are fast rotators.

Recent surveys spanning much larger volumes
have been able to revisit this issue including the trends with environment.
\citet{Brough17}, \citet{Veale17} and \citet{Greene17} using the SAMI, MASSIVE and MaNGA surveys, respectively,
found that the fraction of slow rotators depends
strongly on stellar mass, with a very weak or no dependence on environment once stellar mass is controlled for 
(see \citealt{Houghton13,DEugenio13} for earlier studies on cluster regions). 
They found that the original environmental dependence reported in ATLAS$^{\rm 3D}$ \citep{Emsellem11} was fully
accounted for by massive galaxies preferentially living in denser environments.
Interestingly, the three surveys reached the same conclusion despite the very different
environments and mass ranges studied.
\citet{Brough17} focused on cluster galaxies only, while \citet{Greene17}
covered a much wider halo mass range, $M_{\rm halo}=\left(<10^{12}\,\rm M_{\odot},10^{15}\,\rm M_{\odot}\right)$.
\citet{Veale17} on the other hand make no environmental selection, but only
study galaxies with stellar masses $\gtrsim 10^{11}\,\rm M_{\odot}$.
Note, however, that \citet{Greene17} observed a weak trend for satellite galaxies
to display a slightly higher frequency of slow rotation than centrals at fixed stellar mass, but this 
trend is not significant. Thus, the question of whether there is an environmental effect
on the incidence of slow rotation or not, and in which regimes it is more likely to be significant,
remains unanswered.

The early results from SAURON and ATLAS$^{\rm 3D}$ prompted
a wealth of simulations and theoretical work.
\citet{Jesseit09}, \citet{Bois11} and \citet{Naab14}, based on
simulations of modest numbers of galaxies, found that
the formation paths of slow and fast rotators can be highly varied. \citet{Naab14} showed that slow rotators
could be formed as a result of wet {or dry} major mergers, or by dry minor mergers. In the case of
wet mergers, the remnant can be either a fast or a slow rotators,
or even a disk (e.g.
\citealt{Springel00}; \citealt{Cox06};
\citealt{Robertson06}; \citealt{Johansson09}; \citealt{DiMatteo09};
\citealt{Peirani10}; \citealt{Lotz10};
\citealt{Naab14}; \citealt{Moreno15}).
\citet{Sparre17}, however, found
that galaxy remnants of major mergers can easily evolve into star-forming disk galaxies unless sufficiently strong feedback 
is present to prevent the disk regrowth.
Similarly, \citet{Moster11} concluded that even a dry merger remnant can become a fast rotator
if the surrounding gaseous halo continues to cool down, fuelling the central galaxy and leading to disk regrowth.
 {\citet{Naab14} and \citet{Li18} show that the shapes and the velocity anisotropies of galaxies can provide 
unique clues that may help disentangle the merger history of galaxies.} 

Although valuable insight can be gained from the idealised and cosmological zoom-in simulations
above, they struggle to shed light into the effect of environment and
in having an unbiased representation of different formation pathways. The latter
comes naturally from large, cosmological hydrodynamical simulations, which have the
ability to simultaneously follow the evolution of tens of thousands of galaxies
in a very wide range of environments.
Recently, there has been a major breakthrough
in the capability of cosmological hydrodynamical simulations to produce realistic galaxy populations. 
This has been achieved thanks to improved subgrid models for unresolved
feedback processes, the calibration of subgrid feedback parameters to
match key observables, and the ability to run large cosmological volumes
with sub-kpc resolution. 
Examples of these simulations include \eagle\ \citep{Schaye14}, Illustris \citep{Vogelsberger14} and
its successor Illustris-TNG \citep{Pillepich17}, and
Horizon-AGN \citep{Dubois14}. 

The simulations above reproduce, with various degrees
of success, the morphological diversity
of galaxies observed in the local Universe,
the galaxy colour bimodality, the SFR-stellar mass relation, the stellar mass function
and the cosmic SFR density evolution (e.g. \citealt{Furlong14}; \citealt{Genel14}; \citealt{Trayford15,Trayford16};
\citealt{Synder15}; \citealt{Dubois16}; \citealt{Nelson17}).
Recently, \citet{Penoyre17} analysed the formation path of thousands of elliptical
galaxies in Illustris 
and concluded that major mergers were the most important formation path of slow rotators.
Surprisingly, \citet{Penoyre17}
found no significant difference between the effect of dry vs. wet mergers on the spin of galaxies, in 
contradiction with the work of \citet{Naab14} on cosmological zooms.
{\citet{Li18}, also on the Illustris simulation, showed that the orbital parameters of the merger 
can affect the rotation of the remnant galaxy, with circular orbits preferentially 
producing fast rotators (see also \citealt{Lagos17}).}

In this paper we use the \eagle\ and \Hydrangea\ simulations with the aim
of exploring how the frequency of slow rotators depend on mass and environment. 
\eagle\ simulated a box of $100\rm (\rm cMpc)^3$, while 
\Hydrangea\ is a suite of $24$ cosmological zoom-in simulations of galaxy clusters and their environments \citep{Bahe17}, 
which is part of the larger Cluster-EAGLE project \citep{Barnes17}. The latter consists of 
$30$ galaxy clusters ($6$ more than \Hydrangea). The advantage of using 
\Hydrangea\ here is that it resolves a larger Lagrangian region of $10\,\rm r_{\rm 200}$ for each cluster (as oppose to 
$5\,\rm r_{\rm 200}$ in Cluster-EAGLE), allowing 
us to study groups around clusters.
Together \eagle\ and \Hydrangea\ span the halo mass range $10^{11}\,\rm M_{\odot}-10^{15.3}\,\rm M_{\odot}$ and provide large statistics.
Given this wide dynamic range, we expect our simulations to be able to reveal an environmental dependence 
of the fraction of slow rotators if any is present.
Our aim is to connect these dependencies with the different formation paths of slow rotators and
to disentangle nurture vs. nature in their formation.

\eagle\ is an ideal testbed for our analysis, as it has been shown to reproduce the
size-stellar mass relation \citep{Furlong15,Katsianis17} and the specific angular momentum-stellar mass
relation (\citealt{Lagos16b}; \citealt{Swinbank17}) throughout time, both of which reflect
the ability of the simulation to reproduce structural and dynamical properties of galaxies.
In addition, \eagle\ reproduces very well the evolution of SFR properties of galaxies \citep{Furlong14},
colours \citep{Trayford15}, the gas contents of galaxies \citep{Bahe15,Lagos15,Lagos15b,Crain16}, 
and produces both a blue cloud of predominantly disky galaxies, and a red sequence of mostly 
elliptical galaxies \citep{Correa17}.

This paper is organised as follows. In $\S$~\ref{EagleSec} we briefly describe the \eagle\ simulation suite and 
introduce the IFU-like cubes and the kinematic properties we measure in the simulated galaxies.
$\S$~\ref{VSselec} presents an analysis of the kinematic properties of simulated galaxies
at $z=0$ and the dependence on mass, environment and morphology. Here, we also present
a thorough comparison with observations. In $\S$~\ref{physicalorigin} we study the physical
origin of slow rotators in \eagle\ by connecting kinematics with the formation history
of galaxies. We present a discussion of our results and our main conclusions in $\S$~\ref{conclusions}.
Finally, Appendix~\ref{ConvTests} presents our convergence studies.

\section{The EAGLE simulation}\label{EagleSec}

The \eagle\ simulation suite
(described in detail by \citealt{Schaye14}, hereafter
S15, and \citealt{Crain15}, hereafter C15) consists of a large number of cosmological
hydrodynamic simulations with different resolutions, cosmological volumes and subgrid models,
adopting the \citet{Planck14} cosmological parameters.
A major aspect of the \eagle\ project is the use of
state-of-the-art sub-grid models, which include: (i) radiative cooling and
photoheating \citep{Wiersma09b}, (ii) star formation \citep{Schaye08}, (iii) stellar evolution and chemical enrichment \citep{Wiersma09},
(iv) stellar feedback \citep{DallaVecchia12}, and (v) black hole growth and active galactic nucleus (AGN) feedback 
\citep{Rosas-Guevara13}. S15 introduced a reference model, within which the parameters of the
sub-grid models governing energy feedback from stars and accreting black holes (BHs) were calibrated to ensure a
good match to the $z=0.1$ galaxy stellar mass function and
the sizes of present-day disk galaxies.
Table~\ref{TableSimus} summarises the parameters
of the simulation used in this work.
Throughout the text we use pkpc to denote proper kiloparsecs and
cMpc to denote comoving megaparsecs.

\begin{table}
\begin{center}
  \caption{Features of the \eagle\ Ref-L100N1504 and Ref-L050N752 and simulations used in this paper. The rows list:
    (1) initial particle masses of gas and (2) dark
    matter, (3) comoving Plummer-equivalent gravitational
    softening length, and (4) maximum physical 
    gravitational softening length. Units are indicated in each row. \eagle\
    adopts (3) as the softening length at $z\ge 2.8$, and (4) at $z<2.8$. These two simulations 
    have volumes of side $L=100$ and $50$~$\rm cMpc^3$, respectively.}\label{TableSimus}
\begin{tabular}{l l l l}
\\[3pt]
\hline
& Property & Units & Value \\
\hline
(1)& gas particle mass & $[\rm M_{\odot}]$ & $1.81\times 10^6$\\
(2)& DM particle mass & $[\rm M_{\odot}]$ & $9.7\times 10^6$\\
(3)& Softening length & $[\rm ckpc]$ & $2.66$\\
(4)& max. gravitational softening & $[\rm pkpc]$& $0.7$ \\
\end{tabular}
\end{center}
\end{table}


In addition to the \eagle\ suite, we also analyse the \Hydrangea\ suite presented in 
\citet{Bahe17}. This suite consists of $24$ cosmological zoom-in simulations of galaxy clusters and their large scale environments 
in the halo mass range $M_{\rm 200}=10^{14}-10^{15.4}\,\rm M_{\odot}$, 
with $M_{\rm 200}$ denoting the total mass within a sphere of radius $r_{\rm 200}$, 
within which the average density
equals $200$ times the critical density.
These clusters were simulated with the same \eagle\ reference model, 
but with a higher temperature to which AGN heat nearby gas particles, $\Delta\,T_{\rm AGN}$, and a higher viscosity parameter, $C_{\rm visc}$, 
{ that controls the effect of angular momentum on
black hole gas accretion}. 
The reference \eagle\ model adopted $\Delta\,T_{\rm AGN}=10^{8.5}\rm K$ and $C_{\rm visc}=2\pi$, while 
\Hydrangea\ adopted $\Delta\,T_{\rm AGN}=10^{9}\rm K$ and $C_{\rm visc}=2\pi\times 10^2$ 
(this model is referred to as AGNdT9 in S15; see their Table~$3$). 
{S15 compared the stellar mass function and size-mass relation at $z=0$ of these two models (their Figs.~$9$ and $11$), 
and showed that they agree to better than $10$\% and $20$\%, respectively.} 
The \Hydrangea\ outputs were analysed with the same tools employed in \eagle, 
and described above. In Appendix~\ref{modelcomp} we compare 
the AGNdT9 and reference models 
on the same box, number of particles and initial conditions, and show 
that AGNdT9 tends to produce a very similar number of slow rotators 
at $10^{9.5}\,\rm M_{\odot} \gtrsim M_{\rm stars} \gtrsim 10^{11}\,\rm M_{\odot}$ compared to the reference \eagle\ model ($\approx 9$\%).

{Throughout the text we will refer to `central' and `satellite' galaxies, where 
the central corresponds to the galaxy hosted by the main subhalo of a Friends-of-Friends halo, 
while other subhalos within the group host satellite galaxies \citep{Qu17}.}
{\citet{Lagos16b} showed in a study of the specific angular momentum evolution of galaxies in \eagle, 
that an appropriate stellar mass cut above which galaxies have angular momentum profiles 
converged is $M_{\rm stars}>5\times 10^{9}\,\rm M_{\odot}$. Thus, we adopt that threshold in this work 
(see Appendix~\ref{rescovapp} for a convergence study). {\eagle\ and \Hydrangea\ have $5,587$ and $10,771$ galaxies, respectively, at $z=0$ above this mass 
threshold, which compose the sample used in this work.}  }

\subsection{Kinematic measurements}\label{kinematicmeasurements}

In this paper we measure the $r$-band luminosity-weighted 
line-of-sight velocity, velocity dispersion, stellar spin parameter $\lambda_{\rm R}$, and ellipticity of 
all galaxies in \eagle\ in the simulations presented in Table~\ref{TableSimus} and the \Hydrangea\ 
clusters. We describe our procedure below.

We first construct the stellar kinematic maps for each galaxy by projecting them onto 
a $2$-dimensional plane. We use two orientations: an edge-on view, in which the stellar spin 
is oriented along the $y$-axis of the image, and a random view, in which the line-of-sight
is along the $z$-axis of the simulated box. 
We bin this $2$-dimensional image onto pixels of width $w$ and construct a $r$-band luminosity-weighted 
velocity distribution for each bin, using the centre of potential of the galaxy as 
the rest frame. 
We adopted $w=1.5\,\rm pkpc$ (approximately twice the softening length of \eagle; see Table~\ref{TableSimus}). 
In Appendix~\ref{convbin} we show that the kinematic 
properties we measure are converged to better than $10$\%. We only 
see significant convergence issues if the bin is chosen to be close to the softening length 
of the simulation {or in galaxies of stellar masses $\approx 5 \times 10^9\,\rm M_{\odot}$ 
when the bin is too similar to their $r_{\rm 50}$}. 
The chosen bin of $1.5$~pkpc is very similar to the average spatial resolution 
of SAMI galaxies ($1.6$~pkpc; \citealt{VandeSande17}).

\begin{figure*}
\begin{center}
\includegraphics[trim=2mm 2mm 0mm -3mm, clip,width=0.7799\textwidth]{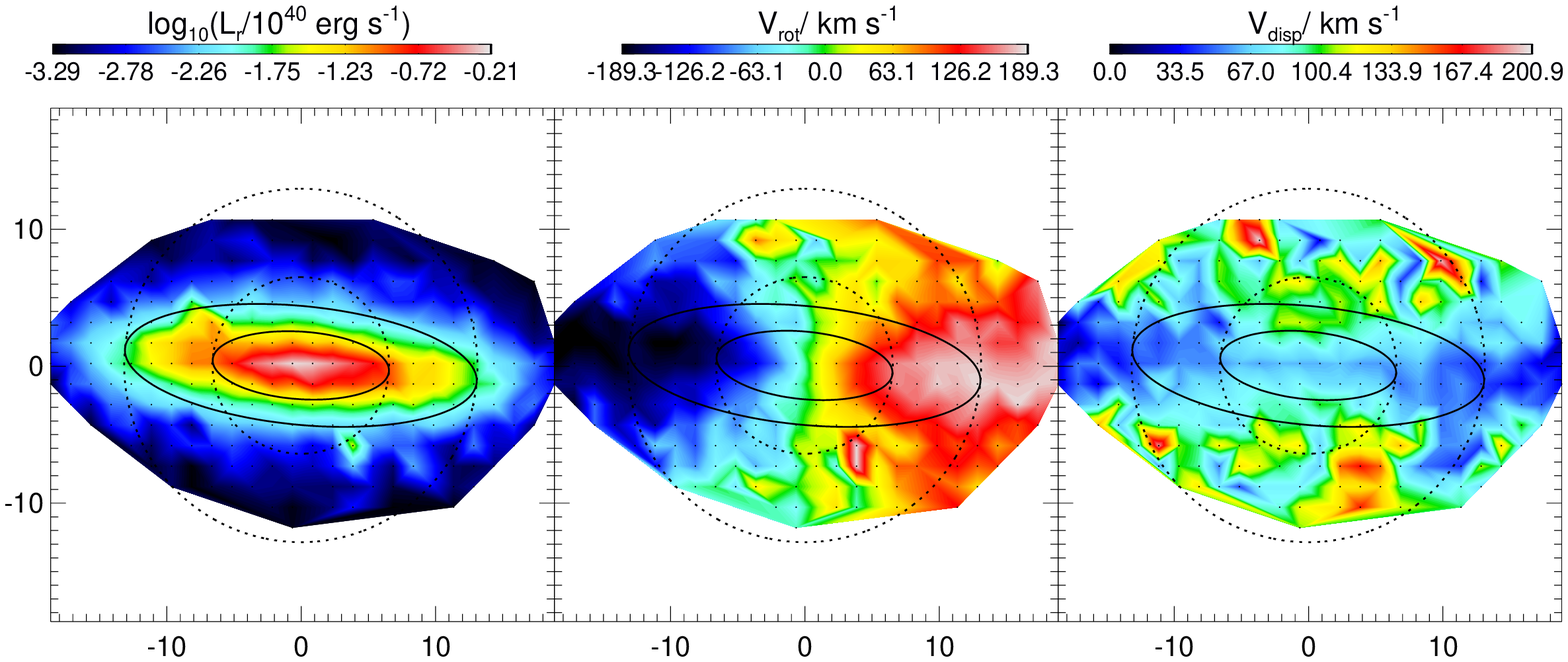}
\includegraphics[trim=2mm 2mm 0mm 2mm, clip,width=0.7799\textwidth]{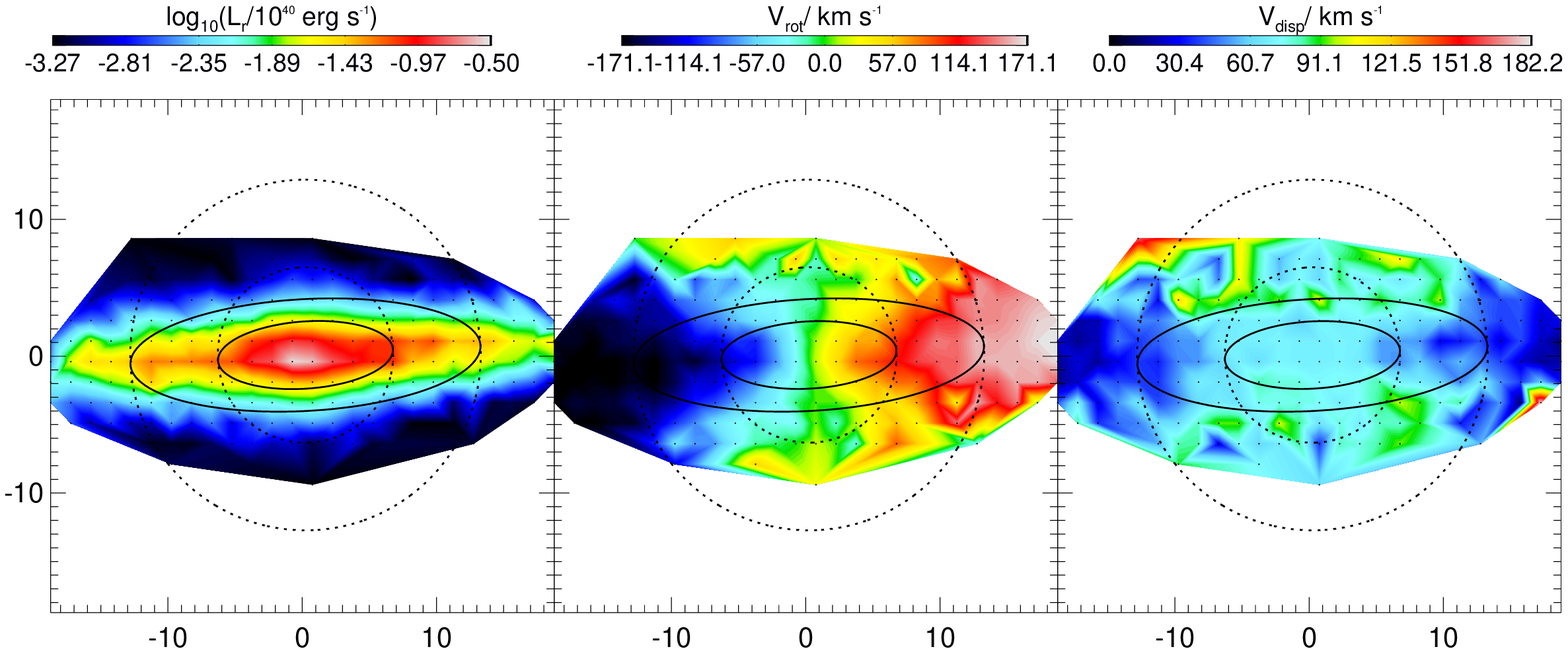}
\includegraphics[trim=2mm 2mm 0mm 2mm, clip,width=0.7799\textwidth]{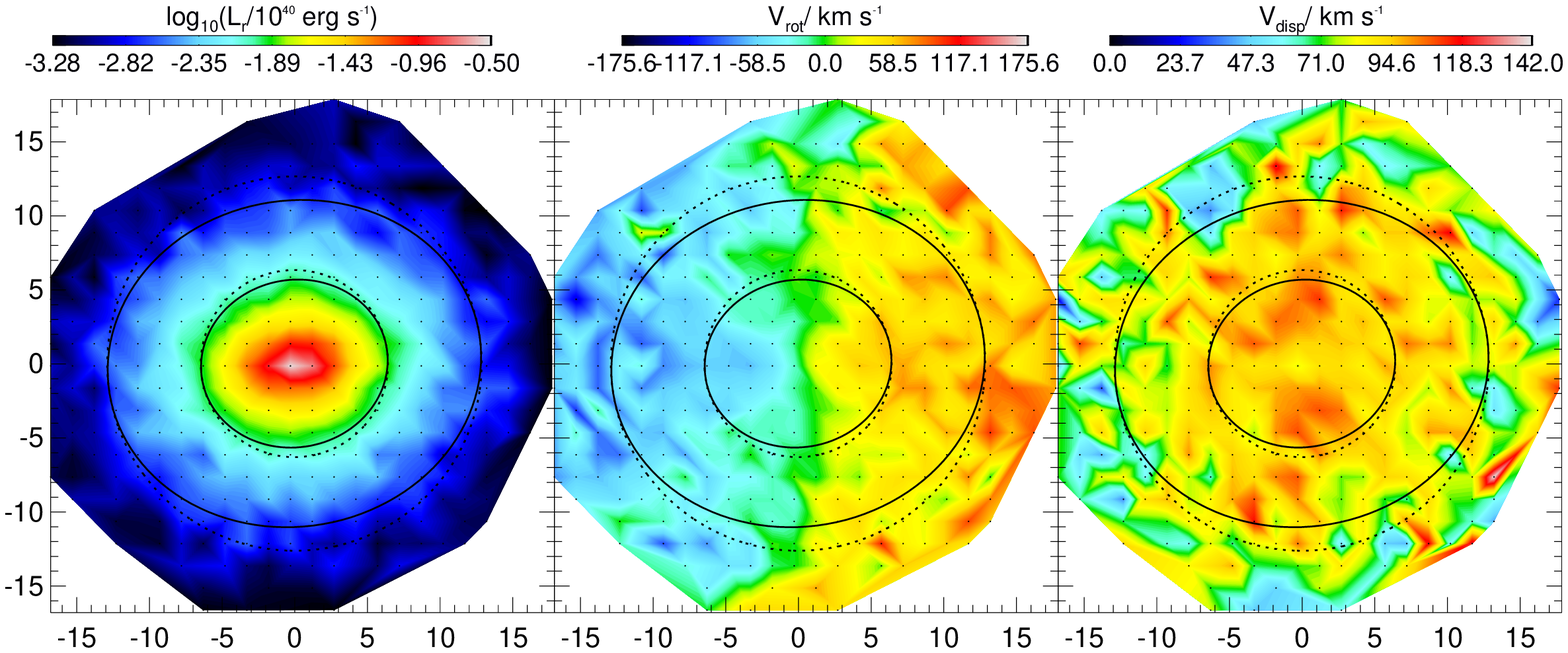}
\includegraphics[trim=2mm 2mm 0mm 2mm, clip,width=0.7799\textwidth]{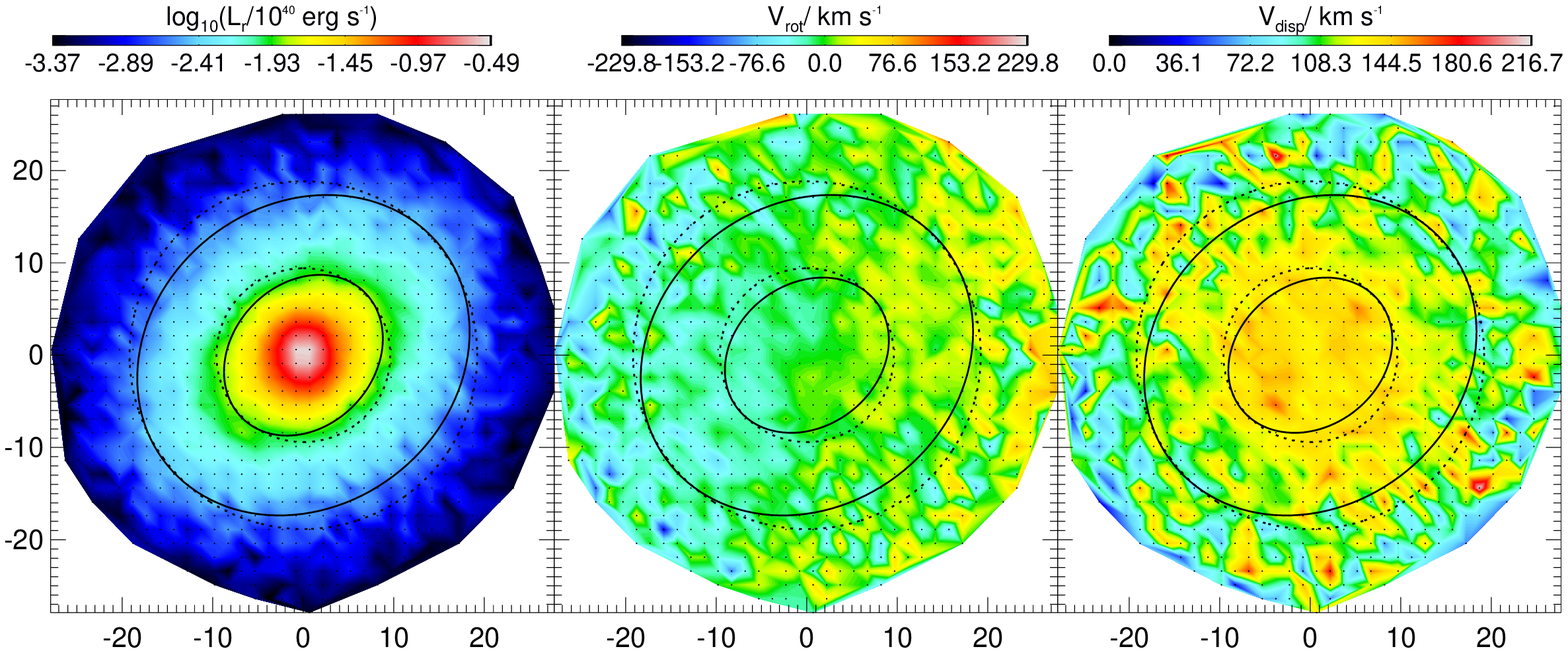}
\caption{Examples of an edge-on view of the $r$-band luminosity (left), rotation (middle) and velocity dispersion (right) fields of $4$ galaxies with 
$2\times 10^{10}\,\rm M_{\odot}<M_{\rm stars}<10^{11}\,\rm M_{\odot}$ at $z=0$ 
in the Ref-L050N752 simulation. Axes show distance from galaxy centre in pkpc.
The top $2$ galaxies have $\rm SFR>1\,\rm M_{\odot}\,yr^{-1}$, while the bottom $2$ 
 have $\rm SFR<0.2\,\rm M_{\odot}\,yr^{-1}$. The colour scales are indicated at the top of each panel, and in the case 
of the rotational velocity map, we force the 
range $\left[-V_{\rm max},V_{\rm max}\right]$, where $V_{\rm max}$ is the maximum circular velocity expected for the 
stellar mass of the galaxy given the Tully-Fisher relation measured by \citet{Dutton11}. The physical scale 
of the images is shown along the axes and is in pkpc. Circles show $1$ and $2\, r_{\rm 50}$ of the galaxies, 
while ellipses are constructed using our ellipticity measurements at $1$ and $2\, r_{\rm 50}$ (see Eq.~\ref{ellip}). 
From top to bottom, the values of $\epsilon_{\rm r_{50}}$ are $0.6$, $0.65$, $0.12$ and $0.19$, respectively.}
\label{examples}
\end{center}
\end{figure*}

We fit a Gaussian to the $r$-band luminosity line-of-sight velocity distribution of each pixel, and define 
the rotational velocity as the velocity at which the Gaussian peaks, and the velocity dispersion 
as the square root of the variance. This procedure closely mimics the measurements
performed in integral field spectroscopic (IFS) surveys, such as ATLAS$^{3D}$ \citep{Cappellari11} and 
SAMI \citep{VandeSande17}. The result of this procedure is shown in Fig.~\ref{examples} for $4$ 
relatively massive galaxies in the Ref-L050N752 simulation, $2$ star-forming and $2$ passive, oriented edge-on. 
For this visualization we use the {\sc kinemetry} package of \citet{Krajnovic06}, 
and for the colour scale of the rotational velocity maps we adopt the range $\left[-V_{\rm max},V_{\rm max}\right]$. 
Here, $V_{\rm max}$ is the maximum circular velocity expected for the
stellar mass of the galaxy assuming the Tully-Fisher relation measured by \citet{Dutton11}. 
The purpose of this colour scheme is to make slow rotation visually evident. In general, we find that 
at fixed stellar mass, passive galaxies tend to be rounder and more slowly rotating than 
star-forming galaxies. We will come back to this in $\S$~\ref{VSselec}.

We construct velocity and luminosity maps, such as those in Fig.~\ref{examples}, for all galaxies 
in the simulations of Table~\ref{TableSimus} and in the \Hydrangea\ cluster suite at $z=0$. 
From these maps we calculate the $r$-band luminosity-weighted 
 spin parameter, $\lambda_{\rm R}$ at radii $0.5, 1, 1.5, 2,\times \rm r_{\rm 50}$, with $r_{\rm 50}$
being the projected half-stellar mass radius.
The ellipticities, $\epsilon$, {are calculated in the same apertures from the projected positions 
of particles} following \citet{Cappellari07},
%

\begin{equation}
	\epsilon = 1-\sqrt{\frac{a^2}{b^2}}, 
\label{ellip}
\end{equation}

\noindent where, 

\begin{eqnarray}
a^2 &=& \frac{\bar{x}^2+\bar{y}^2}{2} + \sqrt{\left(\frac{\bar{x}^2-\bar{y}^2}{2}\right)^2+\bar{xy}},\nonumber \\
b^2 &=& \frac{\bar{x}^2+\bar{y}^2}{2} - \sqrt{\left(\frac{\bar{x}^2-\bar{y}^2}{2}\right)^2+\bar{xy}},
\label{ellip2}
\end{eqnarray}

\noindent and,

\begin{equation}
\bar{x}^2 = \frac{\sum_i L_i\,x^2_i}{\sum_i L_i},\,\bar{y}^2 = \frac{\sum_i L_i\,y^2_i}{\sum_i L_i},\,\bar{xy} = \frac{\sum_i L_i\,x_i\,y_i}{\sum_i L_i}.
\label{ellip3}
\end{equation}

\noindent Here, $i$ corresponds to the stellar particles inside the aperture in which we wish to measure 
$\epsilon$, $L_{i}$ is the $r$-band luminosity of {the particle, $(x_i,y_i)$ are their $x$ and $y$ 
positions in the projected map}.
This measurement of $\epsilon$ is equivalent to diagonalizing the inertia tensor of the galaxy's 
luminosity surface density. We also calculate the position angle of the major axis of the galaxy (measured 
counter clockwise from $y=0$) as 

\begin{equation}
	\theta_{\rm PA} = {\rm atan}\left(\frac{2\,\bar{xy}}{\bar{x}^2-\bar{y}^2}\right).
	\label{theta}
\end{equation}

\noindent Examples of the values of $\epsilon$ 
obtained via this method are shown in Fig.~\ref{examples}. We then calculate 
$\lambda_{\rm R}$ as 

\begin{eqnarray}
\lambda_{\rm R} &=& \frac{\sum_j L_j\, r_j |V_j|}{\sum_j L_j\, r_j \sqrt{V^2_j+\sigma^2_j}},\label{lambdaR}
\end{eqnarray}

\noindent where $V_j$ and $\sigma_j$ are the $r$-band luminosity-weighted 
line-of-sight mean and standard deviation velocities in the pixel $j$ {of the velocity maps} calculated as described 
above, and $r_j$ is the distance from the centre of the galaxy to the pixel (i.e. the circular radius). 
As in \citet{Emsellem11}, 
to measure these quantities within $r$, we include only pixels enclosed by the ellipse 
of major axis $r$, ellipticity $\epsilon(r)$ and position angle $\theta_{\rm PA}$(r).

IFS surveys typically compare $\epsilon$ and $\lambda_{\rm R}$ 
measured within the same aperture (typically an effective radius; e.g. \citealt{Emsellem11} and 
 \citealt{VandeSande17}). We follow this and 
compare $\lambda_{\rm R}$ and $\epsilon$ measured within $r_{\rm 50}$, and refer to these as $\lambda_{\rm r_{50}}$ and 
$\epsilon_{r_{50}}$, respectively, 
unless otherwise stated.

\subsection{Galaxy mergers}\label{galmergerssec}

We use the merger trees available in the \eagle\ database \citep{McAlpine15} to identify galaxy mergers (see \citealt{Qu17} 
for details on how these trees are constructed).
Galaxies that went through mergers have more than one progenitor, and 
we track the most massive progenitor to compare their 
kinematic properties with that of the merger remnant.
The trees used here connect $29$ epochs, with time span between snapshots ranging from $\approx 0.3$~Gyr to $\approx 1$~Gyr.
\citet{Lagos17} showed that these timescales are appropriate to study the effect of galaxy mergers 
on the specific angular momentum of galaxies, as mergers roughly take that time to settle. 

We split mergers into major and minor mergers.
The former are those with a stellar mass ratio between the secondary and the primary galaxy $\ge 0.3$,
while minor mergers have a mass ratio between $0.1$ and $0.3$. Lower mass ratios are 
 classified as smooth accretion \citep{Crain16}. In addition, and following \citet{Lagos17}, 
 we split mergers into gas-rich (wet) and gas-poor (dry) based 
on the neutral gas (atomic plus molecular) to stellar mass ratio of the merger:

\begin{equation}
R_{\rm gas,merger} \equiv \frac{M^{\rm s}_{\rm neutral}+M^{\rm p}_{\rm neutral}}{M^{\rm s}_{\rm stars}+M^{\rm p}_{\rm stars}},
\label{fgasmerg}
\end{equation}

\noindent where $M^{\rm s}_{\rm neutral}$ and $M^{\rm p}_{\rm neutral}$ 
are the neutral gas masses of the secondary and primary galaxies,
respectively,
while $M^{\rm s}_{\rm stars}$ and $M^{\rm p}_{\rm stars}$ are the corresponding stellar masses.
Here we classify mergers with $R_{\rm gas,mergers} \le 0.1$ as dry, and the complement as wet.
For dry mergers, the average $R_{\rm gas,merger}$ is $\approx 0.02$.

{We calculate the orbital specific angular momentum of the merger, $j_{\rm orbital}$,  
as $j_{\rm orbital}=|\vec{r}\times \vec{v}|$, 
with $\vec{r}$ and $\vec{v}$ being the position and velocity vectors, respectively, of the secondary galaxy in the 
rest frame of the primary.}

Masses are measured within an aperture of $30$~pkpc. 
The fraction of atomic and molecular gas in gas particles are calculated in post-processing
following \citet{Rahmati13} and \citet{Lagos15}.

\section{Kinematic properties of \eagle\ galaxies}\label{VSselec}

\begin{figure*}
\begin{center}
\includegraphics[trim=0mm 0mm 0mm 0mm, clip,width=0.73\textwidth]{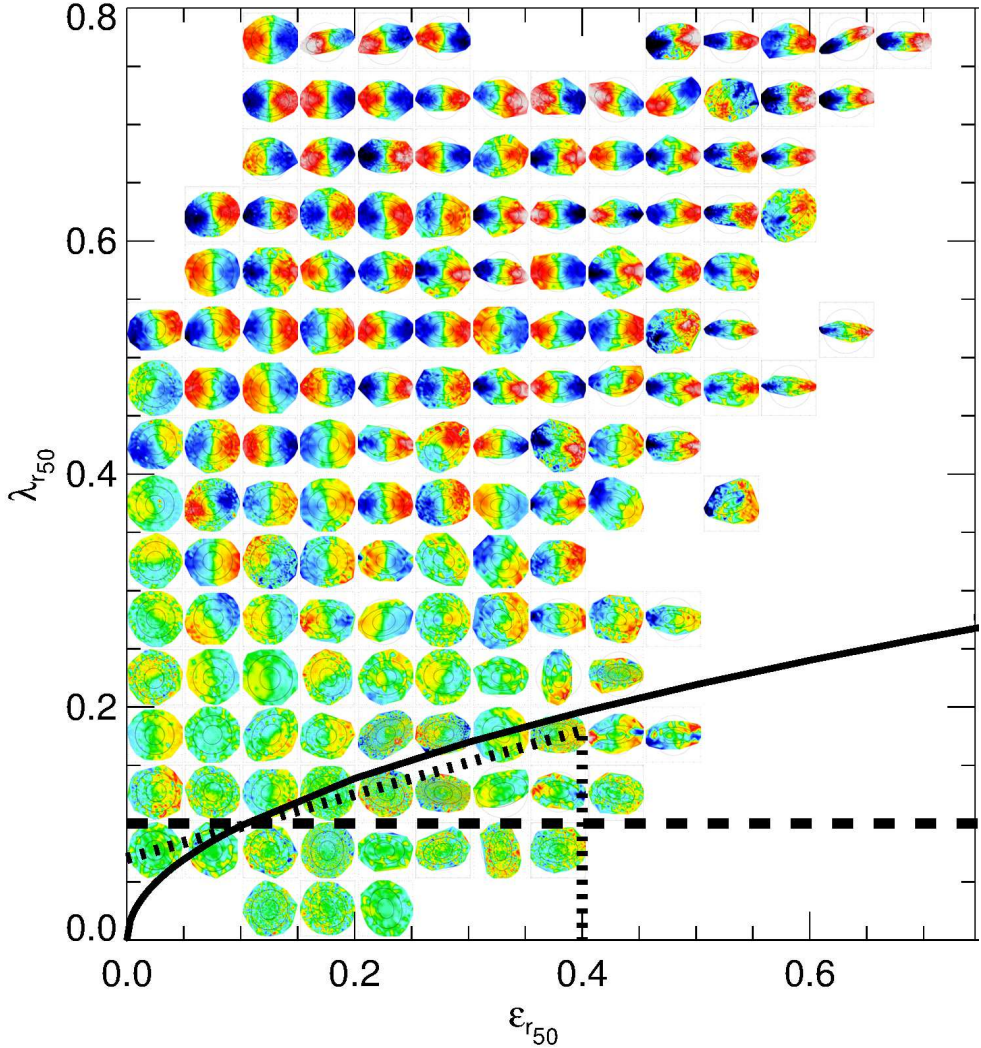}
\caption{Rotational velocity field of randomly selected galaxies in the $\lambda_{r_{50}}$-$\epsilon_{\rm r_{50}}$ plane from the Ref-L050N752 simulation. 
Galaxies here are randomly oriented. 
The colour scales of the maps and circles/ellipses are as in Fig.~\ref{examples}. 
Lines show the classification of slow rotators from \citet{Emsellem07}, \citet{Emsellem11} and
\citet{Cappellari16}, as dashed, solid and dotted lines, respectively.}
\label{VSStellarMassColors}
\end{center}
\end{figure*}

\begin{figure}
\begin{center}
\includegraphics[trim=4mm 5mm 0mm 0mm, clip,width=0.5\textwidth]{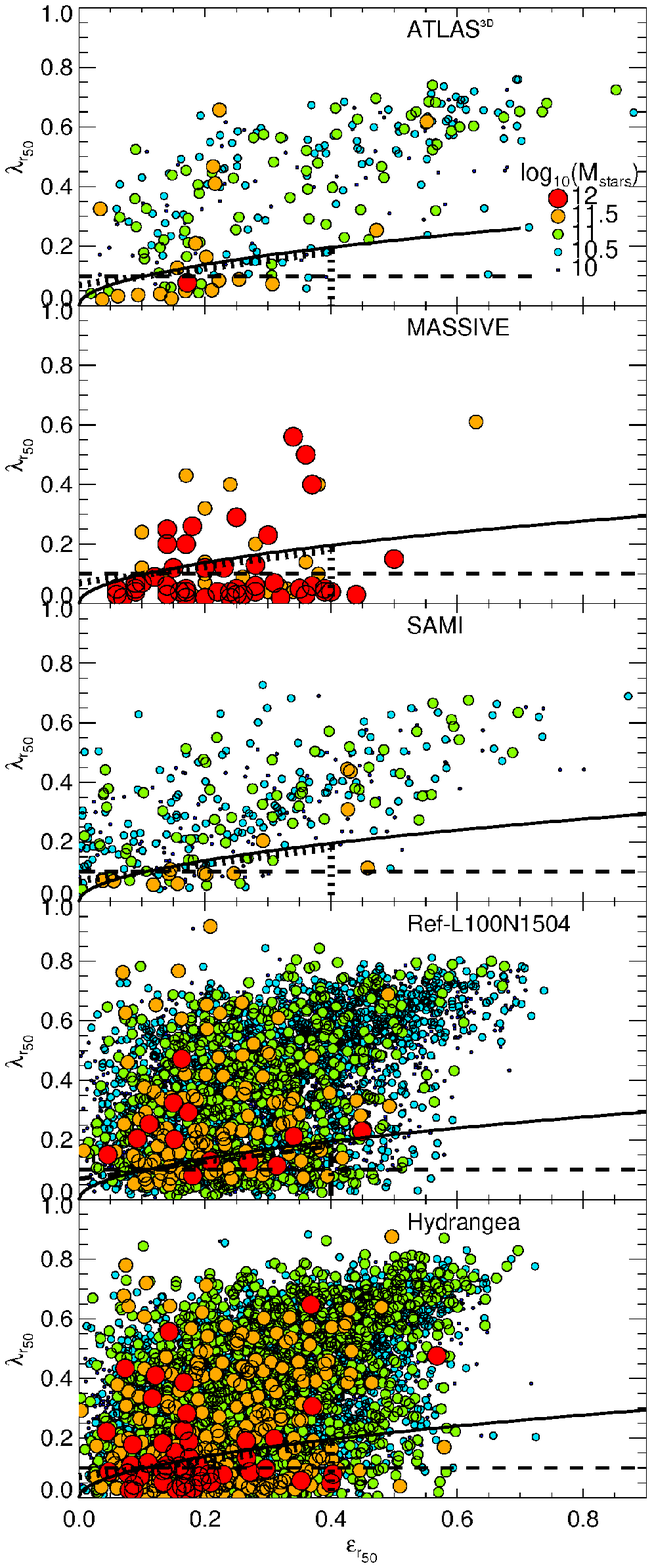}
\caption{$\lambda_{\rm r_{50}}$ as a function of $\epsilon_{\rm r_{50}}$ for galaxies in the ATLAS$^{\rm 3D}$ (\citealt{Emsellem11}; top panel), 
MASSIVE (\citealt{Veale17b}; second panel) and 
SAMI (\citealt{VandeSande17}; third panel) surveys,
and for the simulations Ref-L100N1504 (fourth panel) and 
\Hydrangea\ (bottom panel).
Galaxies in the two simulations are randomly oriented. 
Lines show the classification of slow and fast rotators from \citet{Emsellem07}, \citet{Emsellem11} and 
\citet{Cappellari16}, as dashed, solid and dotted lines, respectively.
Sizes and colours of the symbols correspond to different stellar masses, as labelled in the top panel.} 
\label{LambdaREllip}
\end{center}
\end{figure}

We visually inspect the kinematic morphology of galaxies in the $\lambda_{r_{50}}$-$\epsilon$ plane, which has been 
proposed by \citet{Emsellem07} as an effective way of distinguishing slow and fast rotators. Fig.~\ref{VSStellarMassColors}
shows the rotational velocity maps of randomly selected galaxies in bins 
of $\lambda_{r_{50}}$ and $\epsilon$. We construct the maps as in Fig.~\ref{examples}. Lines indicate different ways 
of defining slow rotators from the literature. There is an evident transition at around $\lambda_{r_{50}}\approx 0.2$ 
below which galaxies {appear} deficient in rotation. {By inspecting 
the edge-on oriented velocity maps of galaxies that are classified as slow rotators in \eagle, 
we confirm their deficient rotation out to $3\,r_{50}$.}  If \eagle\ galaxies are
 a good representation of real ones, this would mean that slow rotators 
would be classified as such even if we had kinematics extending out to much larger radii than available (typically kinematics is available
only at $r<r_{50}$). On the other hand, galaxies with $0.2 \lesssim \lambda_{r_{50}}\lesssim 0.4$ reach their expected rotational velocity at $2-3\,r_{50}$, 
while galaxies with $\lambda_{r_{50}}\gtrsim 0.4$ reach it by $\approx r_{50}$. Fig.~\ref{VSStellarMassColors} indicates that 
 $\lambda_{r_{50}}$ is a good proxy for the kinematic structure of galaxies, as suggested by \citet{Emsellem07,Emsellem11}.

In Fig.~\ref{LambdaREllip} we visually compare the positions of galaxies in the $\lambda_{r_{50}}$-$\epsilon$ plane 
in the Ref-L100N1504 and \Hydrangea\ simulations with those of the observational surveys ATLAS$^{\rm 3D}$ (\citealt{Emsellem11}), 
MASSIVE \citep{Veale17b} and SAMI (\citealt{VandeSande17}). The former two are volume-limited surveys of early-type galaxies, 
while SAMI is a stellar mass selected survey, 
thus including both late and early types. 
Since we include all galaxies with $M_{\rm stars}>5\times 10^9\,\rm M_{\odot}$ in the simulations, our results may be more comparable to SAMI.
The sizes and colours of the symbols scale with stellar mass, so that the most massive galaxies appear as larger symbols.

SAMI appears to have systematically lower $\lambda_{r_{50}}$ compared to ATLAS$^{\rm 3D}$, which is not surprising 
as the measurements are not performed exactly in the same way. {In ATLAS$^{\rm 3D}$, \citet{Emsellem11} adopted the radial distance to the 
luminosity centre as $r_{i}$ in Eq.~\ref{lambdaR}, while in SAMI, \citet{VandeSande17} adopted the semi major axis of the ellipse that goes 
through the given bin as $r_{i}$ in Eq.~\ref{lambdaR}.}

Our calculation of $\lambda_{r_{50}}$ resembles more closely that in ATLAS$^{\rm 3D}$.
In the three surveys, galaxies with $M_{\rm stars}\gtrsim 10^{11.5}\,\rm M_{\odot}$ (largest symbols in Fig.~\ref{LambdaREllip}) preferentially have low $\lambda_{r_{50}}$, and the same is seen to some extent 
in the \Hydrangea\ simulation, but in the Ref-L100N1504 simulation few massive galaxies below the observational delimitation of slow rotators. 
Compared to MASSIVE (middle panel in Fig.~\ref{LambdaREllip}), it is apparent 
that our simulations do not produce the right fraction of slow rotators at the very massive end. We will come back to this in $\S$~\ref{FracSRSec}.

Both simulations lack the very high ellipticity galaxies, $\epsilon_{r_{50}}\gtrsim 0.75$.
The latter may be due to the subgrid interstellar medium physics included in the simulations, which prevents very flat Milky-Way like disks from forming. 
In \eagle\ a global temperature floor, $T_{\rm eos}(\rho)$, is imposed corresponding 
to a polytropic equation of state $P_{\rm eos}\propto \rho^{\gamma_{\rm eos}}$, normalized 
to $T_{\rm eos}=8,000\,\rm K$ and $n_{\rm H}=0.1\,\rm cm^{-3}$. In addition, a second temperature floor of $8,000\,\rm K$
is imposed on gas with $n_{\rm H}>10^{-5}\,\rm cm^{-3}$, preventing the metal-rich gas from cooling below that threshold. 
This sets a minimum disk height of $\lesssim 1\,\rm kpc$, larger than the Milky-Way or other 
grand-design spiral galaxies, which exhibit scaleheights typically of $\approx 0.4$~kpc \citep{Kregel02}. Thus, it is not surprising that 
very flat galaxies do not exist in \eagle\ or \Hydrangea.
Appendix~\ref{rescovapp} shows that increasing the resolution by a factor of $8$ in mass and $2$ in spatial resolution 
does not significantly change the ellipticity of galaxies, supporting our conclusion.
{The topic of convergence in the formation of elliptical galaxies is contingent.
\citet{Bois10} performed a resolution study of idealised galaxy mergers and concluded that the product of wet mergers 
was resolution dependant, and that their role on the formation of slow rotators may be underestimated in 
simulations such as \eagle. However, more recently \citet{Sparre16,Sparre17} showed in cosmological zooms of galaxy mergers 
that environment and feedback play a decisive role in the fate of the remnant, more so than the resolution.
Our resolution tests show no evidence for convergence issues at the stellar masses we are investigating, on average,  
but we cannot rule out that individual cases may be more affected.}

\begin{figure}
\begin{center}
\includegraphics[trim=0mm 6mm 0mm 0mm, clip,width=0.45\textwidth]{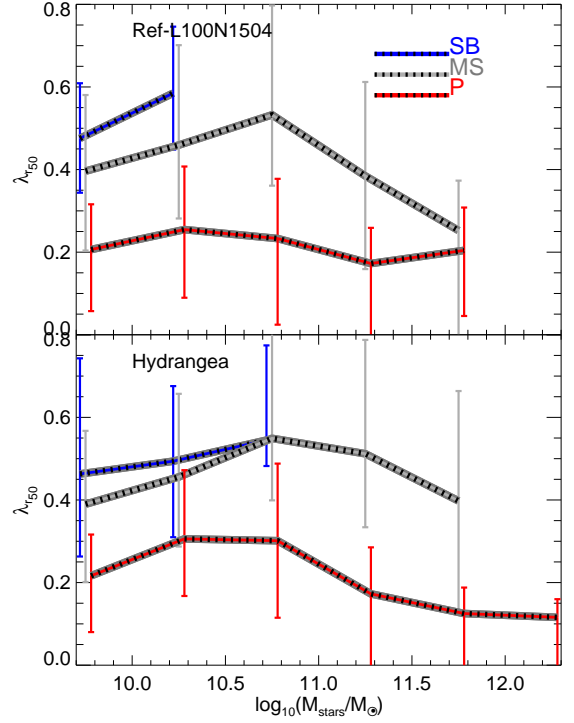}
\caption{$\lambda_{\rm r_{50}}$ as a function of stellar mass for galaxies in the Ref-L100N1504 (top panel) and 
\Hydrangea\ (bottom panel) simulations. Lines and error bars show the median and 
$16^{\rm th}-84^{\rm th}$ percentile ranges, respectively, 
for galaxies that are classifies as starburst (SB), main sequence (MS) 
or passive (P), as labelled. We define the above samples based on $\delta\,\rm MS$: $>4$, $0.1-4$, $<0.1$.
Only bins with $\ge 5$ objects are shown.}
\label{LambdaRMass}
\end{center}
\end{figure}

Fig.~\ref{LambdaRMass} shows $\lambda_{r_{50}}$ as a function of stellar mass for 
galaxies in the Ref-L100N1504 and \Hydrangea\ simulations at $z=0$. We use different symbols to show 
starburst, main sequence and passive galaxies. We define the latter in terms of their specific star formation 
rate, $\rm sSFR=SFR/M_{\rm stars}$, relative to the main sequence at that stellar mass. We 
calculate the latter as in \citet{Furlong14}. In short, the main sequence is calculated as the median 
sSFR of all galaxies that have sSFR$>0.01\rm \,Gyr^{-1}$ in a bin of stellar mass. 
We refer to the this as $\langle sSFR(M)\rangle$. We then calculate 
the sSFR of galaxies relative to the main sequence, 

\begin{equation}
\delta\,\rm MS=\frac{sSFR}{\langle sSFR(M)\rangle}.
\label{MSdef}
\end{equation}

\noindent Starburst (SB), main sequence (MS) and passive galaxies are classified as those with $\delta\,\rm MS\ge 4$, 
$0.1<\delta\,\rm MS<4$ and $\delta\,\rm MS <0.1$, respectively.

In both simulations passive galaxies tend to have a lower $\lambda_{r_{50}}$ than MS galaxies at
fixed stellar mass. SB galaxies have a slightly higher median $\lambda_{r_{50}}$ 
than MS galaxies but the scatter is much larger. In \Hydrangea\ most of the galaxies with 
$M_{\rm stars}\gtrsim 10^{11.7}\,\rm M_{\odot}$ are passive, which is expected given that 
the environments of these simulations are designed to represent the densest in the Universe.
We also see that MS galaxies show a clear peak at $M_{\rm stars}\approx 10^{10.8}\,\rm M_{\odot}$ 
below and above which galaxies display a decrease in $\lambda_{r_{50}}$. This 
peak is also seen in the {\sc FIRE} simulations \citep{El-Badry18}. 
Passive galaxies also exhibit a peak but only in the \Hydrangea\ simulation, which may be due to poor statistics in 
the passive population in the Ref-L100N1504 simulation below that transition mass.

\subsection{The fraction of slow rotators in \eagle}\label{FracSRSec}

\begin{figure}
\begin{center}
\includegraphics[trim=1mm 6mm 0mm 0mm, clip,width=0.48\textwidth]{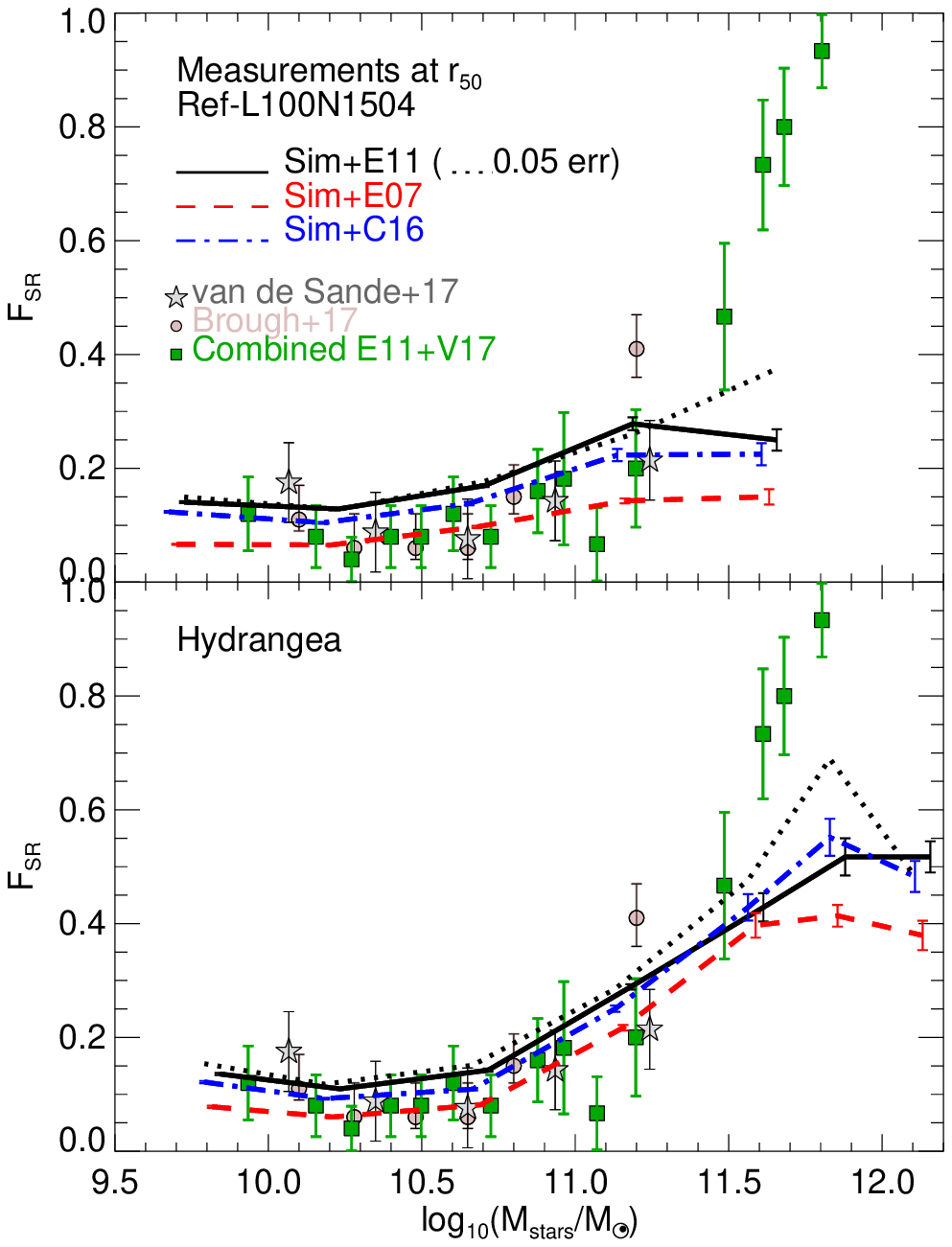}
\caption{The fraction of slow rotators, $\rm F_{\rm SR}$ 
as a function of stellar mass in the Ref-L100N1504 (top panel) and \Hydrangea\ (bottom panel) simulations.
We classified slow rotators using the \citet{Emsellem07} (red dashed line), 
\citet{Emsellem11} (black solid line) and \citet{Cappellari16} (blue dot-dashed line) criteria. 
We also show the observations from the SAMI \citep{VandeSande17}, the SAMI-clusters \citep{Brough17}, MASSIVE \citep{Veale17b} and ATLAS$^{\rm 3D}$ \citep{Emsellem11} surveys. 
The latter two are presented in combination (combined E11+V17) .
{Error bars show $1$ standard deviation calculated with $10$ jackknife resamplings in each stellar mass bin}. 
Dotted lines show $\rm F_{\rm SR}$ adopting the \citet{Emsellem11} criterion 
{after applying a Gaussian error of width $0.05$ 
to the values of $\lambda_{\rm R}$}.}
\label{FSRMass}
\end{center}
\end{figure}

\begin{figure}
\begin{center}
\includegraphics[trim=1mm 6mm 0mm 0mm, clip,width=0.48\textwidth]{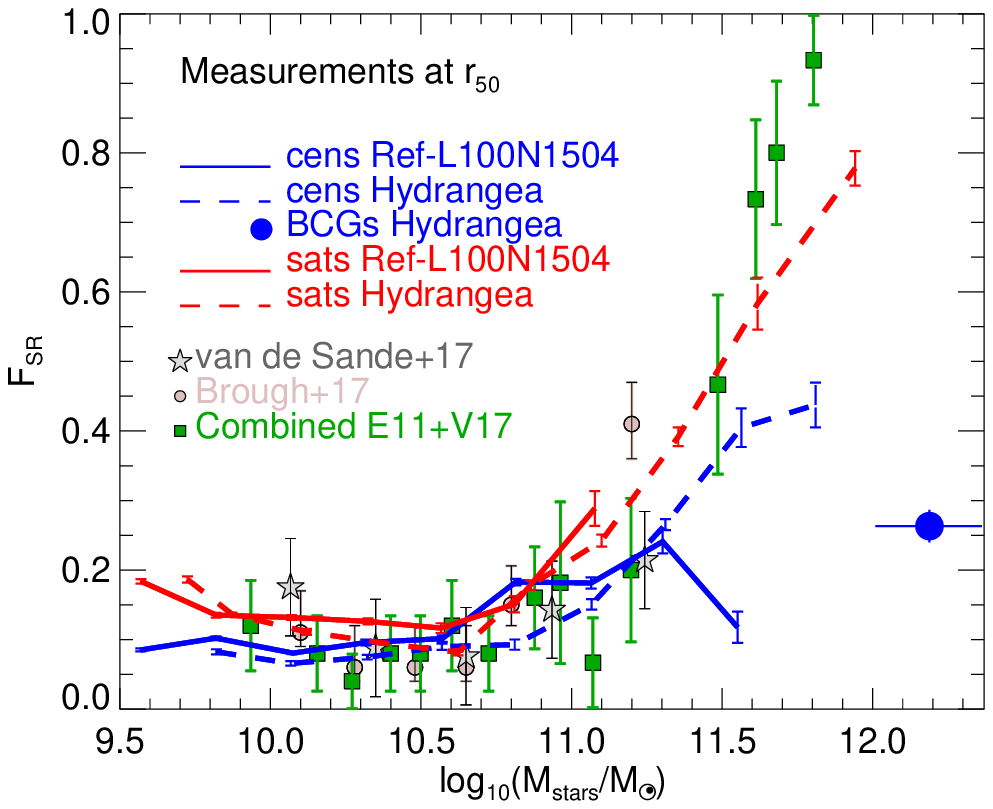}
\caption{$\rm F_{\rm SR}$, as defined by the \citet{Cappellari16} classification, 
as a function of stellar mass, for galaxies at $z=0$ in the Ref-L100N1504 (solid lines)
and \Hydrangea\ (dashed lines) simulations. Central and satellite galaxies are shown in 
blue and red, as labelled. For the \Hydrangea\ simulation we show separately $F_{\rm SR}$ 
for the brightest cluster galaxies as a solid symbol. 
{The horizontal error bar in the latter shows the $25^{\rm th}-75^{\rm th}$ percentile range.}}
\label{FSRMass2}
\end{center}
\end{figure}

Due to the availability of large IFS surveys, there 
has been a lot of recent interest in how the fraction of slow rotators depends on stellar mass 
and environment.
\citet{Veale17}, \citet{Brough17} and \citet{Greene17} found that the fraction of slow rotators depends 
strongly on stellar mass, with a very weak dependence on environment once stellar mass is controlled for.
{Similar results have been reported from the study of galaxy shapes \citep{Pasquali07}.}
The complementarity of \eagle\ and \Hydrangea\ in dynamical mass allows us to explore a very wide range of 
environments and hence to study this question. 

\begin{figure}
\begin{center}
\includegraphics[trim=1mm 6mm 0mm 0mm, clip,width=0.48\textwidth]{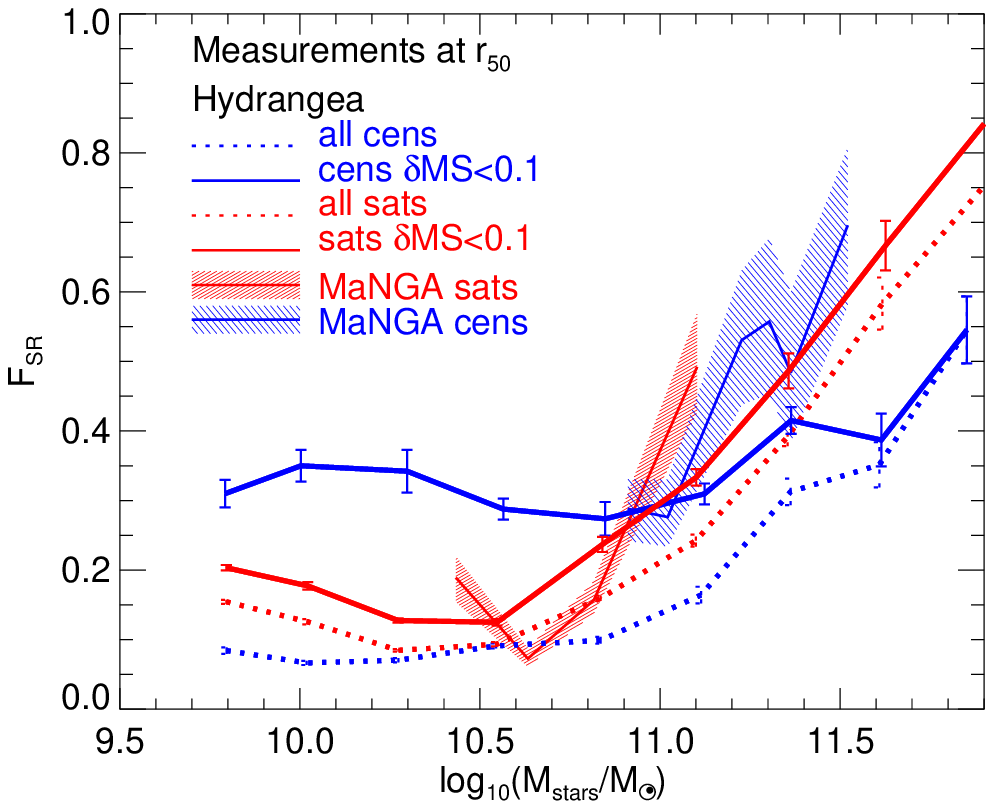}
\caption{The fraction of slow rotators obtained by applying the \citet{Cappellari16} classification 
to the \Hydrangea\ simulations, separating central and satellite galaxies. 
BCGs are not included in this figure. 
Dotted lines show all galaxies in the samples, while 
solid lines show the subsample of galaxies that have a SSFR relative to the MS $\le 0.1$. 
The observations of \citet{Greene17} using MaNGA early-type galaxies are shown as lines with shaded regions indicating the $1\sigma$ scatter.}
\label{FSRMass2Obs}
\end{center}
\end{figure}

Fig.~\ref{FSRMass} shows the fraction of slow rotators, $\rm F_{\rm SR}$, as a function of stellar mass 
at $z=0$ in the Ref-L100N1504 and \Hydrangea\ simulations, using the $3$ definitions of slow rotators 
shown in Fig.~\ref{LambdaREllip}. The two simulations agree well at $M_{\rm stars}\lesssim 10^{10.8}\,\rm M_{\odot}$ within the uncertainties, but there are 
some differences worth noting. 
Both simulations show that there is a clear transition 
at $M_{\rm stars}\approx 10^{11}\,\rm M_{\odot}$ above which $\rm F_{\rm SR}$ starts to raise quickly, 
except for the highest mass bin, in which we see {a flattening or downturn, depending on the criteria adopted 
to classify slow rotators}. In the case of the Ref-L100N1504 simulation,  
this is due to applying the observational classification of slow rotators without 
considering any errors. A small {Gaussian error of width} $0.05$ in $\lambda_{\rm R}$ 
leads to a monotonically rising $\rm F_{\rm SR}$ (dotted lines in Fig.~\ref{FSRMass}).  
\Hydrangea\ displays a downturn at much larger masses ($>10^{11.8}\,\rm M_{\odot}$), and we show in Fig.~\ref{FSRMass2Obs} that this is due to 
 the properties of the brightest cluster galaxies (BCGs) in \Hydrangea. 
{BCGs here are defined as the central galaxy of halos with masses $>10^{14}\,\rm M_{\odot}$. Because 
the \Hydrangea\ suite covers large regions around the $24$ resimulated clusters (out to $10\,\rm r_{\rm 200}$), 
there are in total 34 halos with those masses in the suite, and thus the same number of BCGs.}

In Fig.~\ref{FSRMass} we also show a compilation of observations from the SAMI, ATLAS$^{\rm 3D}$ and 
MASSIVE surveys. Both simulations agree remarkably well with the observations 
at $M_{\rm stars}\lesssim 10^{11.2}\rm \, M_{\odot}$, with some tension arising at $M_{\rm stars}\gtrsim 10^{11.5}\,\rm M_{\odot}$. 
We show below that this is caused by unrealistic properties of our simulated BCGs.
In our simulations, $\rm F_{\rm SR}$ does not rise above $\approx 0.7$ in disagreement with the observations. 
We show later (Fig.~\ref{FSRMass2Obs}) that $\rm F_{\rm SR}$ at $M_{\rm stars}\gtrsim 10^{11.2}\rm \, M_{\odot}$ is very sensitive to environmental effects
and a slightly different preference for satellites over central galaxies can significantly skew 
$\rm F_{\rm SR}$. 

The effect of satellite/central galaxies in the Ref-L100N1504 and \Hydrangea\ simulations 
is shown in Fig.~\ref{FSRMass2}. For clarity, we only show 
the classifications of slow rotators of \citet{Cappellari16}. 
Adopting instead the \citet{Emsellem07,Emsellem11} classifications does not alter the conclusions.
Both simulations show satellite galaxies having a larger $\rm F_{\rm SR}$ 
 compared to centrals (red vs. blue lines), {particularly visible at 
$M_{\rm stars}\gtrsim 10^{10.8}\,\rm M_{\odot}$}. 
However, when selecting only passive galaxies 
(Fig.~\ref{FSRMass2Obs}), centrals have a much 
larger $\rm F_{\rm SR}$ compared to satellites at $M_{\rm stars}\lesssim 10^{11}\,\rm M_{\odot}$.
This is expected as the quenching of central galaxies is typically accompanied by 
morphological transformation, while for satellite galaxies this is not necessary as 
they quench due to the environment they live in (e.g. \citealt{Trayford16,Dubois16}).
The differences between satellites and centrals at fixed stellar mass are significant. We performed 
Kolmogorov-Smirnov tests in narrow bins of stellar mass to quantify how different 
the $\lambda_{\rm r_{\rm 50}}$ distributions between these two populations are and found 
typical $p$~values $<0.05$.

Central galaxies in the \Hydrangea\ simulation show a decrease in $F_{\rm SR}$ in the highest mass bin. This decrease
is significant and is driven by the contribution of BCGs
 (central galaxies of halos with masses $\ge 10^{14}\,\rm M_{\odot}$). 
To make this clearer, we show separately $F_{\rm SR}$ for BCGs in \Hydrangea\ 
as a filled symbol. 
Recently, \citet{Oliva-Altamirano17} analysed a sample of local Universe BCGs and 
found a large fraction of slow rotators, $\approx 50$\%, significantly larger 
than the $26$\% we obtain in \Hydrangea. 
\citet{Bahe17} showed that BCGs in \Hydrangea\ are too massive for their halo mass 
and have some remaining star formation that 
is higher than in observations. Several simulations have shown that continuing star formation 
can efficiently spin galaxies up \citep{Moster11,Naab14,Lagos16b,Penoyre17}, and thus it is not surprising 
that in \Hydrangea\ BCGs are mostly fast rotators. 
It is therefore likely that more efficient feedback at high redshift would not only lead to more realistic 
stellar masses and star formation rates of these BCGs, but also increase their slow rotator fraction (see also \citealt{Barnes17}).

Fig.~\ref{FSRMass2Obs} also shows that satellite galaxies reach an $\rm F_{\rm SR}\gtrsim 0.5$ at 
$M_{\rm stars} \approx 10^{11.5}\,\rm M_{\odot}$ in better agreement with the observations. Since 
both surveys, ATLAS$^{\rm 3D}$ and MASSIVE, are volume-limited, 
$\approx 42$\% of those are satellite galaxies, $24$\% are brightest group/cluster galaxies, and 
the rest are field galaxies.
Thus, it is not surprising that satellite galaxies better follow the results from MASSIVE.
In the Ref-L100N1504 and \Hydrangea\ simulations 
there is a clear environmental effect that becomes apparent when comparing satellites and centrals at fixed stellar mass (Fig.~\ref{FSRMass2}).

\begin{figure}
\begin{center}
\includegraphics[trim=1mm 6mm 0mm 0mm, clip,width=0.48\textwidth]{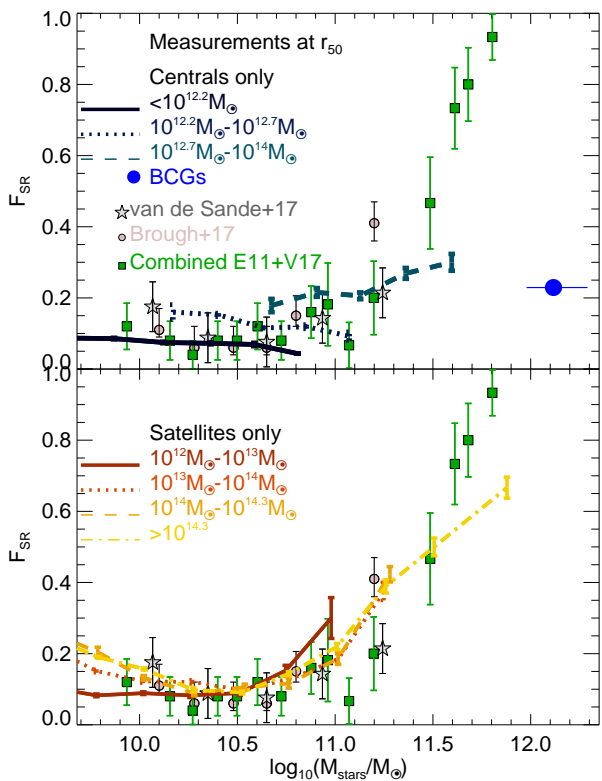}
\caption{{\it Top panel:} $F_{\rm SR}$ as a function of stellar mass for central galaxies at $z=0$
in the combined galaxy sample of the Ref-L100N1504 and \Hydrangea\ simulations. 
Here we only show the slow rotator classification of \citet{Cappellari16}. 
Central galaxies are shown in 4 bins of halo mass: 
$<10^{12.2}\,\rm M_{\odot}$ (solid line), $10^{12.2}\,\rm M_{\odot}-10^{12.8}\,\rm M_{\odot}$ (dotted line), 
$10^{12.8}\,\rm M_{\odot}-10^{14}\,\rm M_{\odot}$ (dashed line) and 
$>10^{14}\,\rm M_{\odot}$ (filled circle; as in Fig.~\ref{FSRMass2Obs}). 
{Error bars show $1$ standard deviation calculated with $10$ jackknife resamplings in each stellar mass bin}.
Observations are shown as symbols, as labelled.
{\it Bottom panel:} as in the top panel but for satellite galaxies. 
Here we adopt halo mass bins of:
$10^{12}\,\rm M_{\odot}-10^{13}\,\rm M_{\odot}$ (solid line), $10^{13}\,\rm M_{\odot}-10^{14}\,\rm M_{\odot}$ (dotted line),
$10^{14}\,\rm M_{\odot}-10^{14.3}\,\rm M_{\odot}$ (dashed line) and
$>10^{14.3}\,\rm M_{\odot}$ (dot-dashed line).
There is a weak but significant systematic effect of $F_{\rm SR}$ increasing with increasing 
halo mass at fixed stellar mass for central galaxies, with no clear trend in the case of satellites.}
\label{FSRMass4}
\end{center}
\end{figure}

\begin{figure}
\begin{center}
\includegraphics[trim=1mm 6mm 0mm 0mm, clip,width=0.48\textwidth]{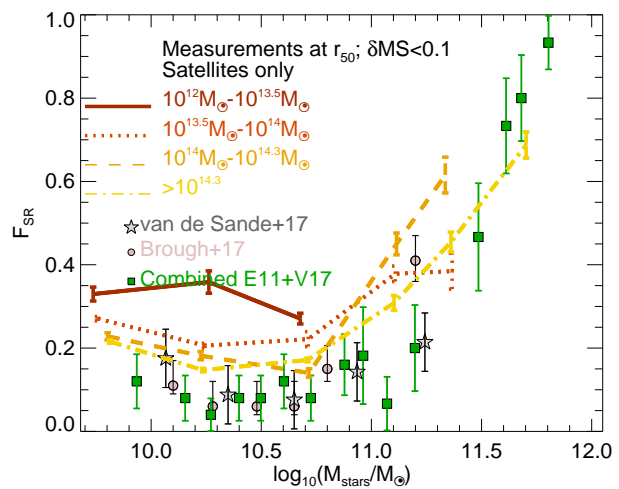}
\caption{As in the bottom panel of Fig~\ref{FSRMass4} but for 
passive satellites (i.e. those with $\delta\,\rm MS<0.1$).}
\label{FSRMass4b}
\end{center}
\end{figure}

Recently \citet{Greene17} found differences between satellite/central early-type galaxies in MaNGA 
of a similar magnitude to the one found in our simulations. 
Their observations are shown as shaded bands in Fig.~\ref{FSRMass2Obs}.
Greene et al. found that satellites are $\approx 20$\% more likely to be slow rotators than 
centrals. They, however, cautioned that 
due to the different spatial coverage and slightly different stellar mass distribution,  
this difference between satellites and centrals is not obviously significant. 
In \eagle\ and \Hydrangea, the $\rm F_{\rm SR}$ differences between these two populations 
are present over the entire mass range, albeit with differences been very small at $M_{\rm stars} < 10^{10.5}\,\rm M_{\odot}$.
Note, however, that the \citet{Greene17} slow rotator fraction is generally higher than 
both the Ref-L100N1504 and \Hydrangea\ simulations and the observations from ATLAS$^{\rm 3D}$, SAMI and MASSIVE.

\citet{Greene17} measured $\lambda_{\rm R}$ out to larger radii than ATLAS$^{\rm 3D}$, SAMI and MASSIVE, 
 and in addition they measured ellipticities from a single S\`ersic index fit to their images.  
\citet{Greene17} adapted the slow rotator classification criteria of \citet{Emsellem11} 
to work with their measurements, and argued that about $10$\% of their galaxies would change classification 
from slow to fast rotators if the measurements were done at $r_{\rm 50}$. Nonetheless, the reported 
$F_{\rm SR}$ is higher by a factor of $\approx 25$\% at least compared to other IFS surveys; hence, there probably are 
other systematic effects that have not yet been taken into account.
Since these authors analysed only early-type galaxies, we also show in Fig.~\ref{FSRMass2Obs} 
the $\rm F_{\rm SR}-M_{\rm stars}$ relation for passive galaxies (those with a $\delta \rm \,MS<0.1$). 
$F_{\rm SR}$ is higher for this subsample but not enough as to agree with \citet{Greene17}. 
Interestingly, in this subsample we find the $F_{\rm SR}$ of centrals being $\approx 2$ times larger than for satellites 
{at $M_{\rm stars}\lesssim 10^{10.7}\rm \,M_{\odot}$}.

We explore the effect of environment on $\rm F_{\rm SR}$ further by studying 
the dependence on halo mass for centrals and satellite galaxies 
in Fig.~\ref{FSRMass4}. Here, we combined the galaxy populations of the 
Ref-L100N1504 and \Hydrangea\ simulations at $z=0$. 
The top panel of Fig.~\ref{FSRMass4} shows central galaxies. 
There is a trend of increasing $\rm F_{\rm SR}$ 
with increasing halo mass, at fixed stellar mass. Since 
stellar and halo mass are well correlated for central galaxies in \eagle\ (\citealt{Schaye14}; \citealt{Guo15}),  
the overlap in stellar mass between the different halo mass bins 
is only partial. 
We quantify the environmental dependence in the stellar mass range where overlap occurs: 
(i) at $M_{\rm stars}= 10^{10.1}-10^{10.8}\,\rm M_{\odot}$,
centrals hosted by halos of masses $10^{12.2}\,\rm M_{\odot}-10^{12.7}\,\rm M_{\odot}$ 
are $2$ times more likely to be slow rotators than centrals in halos of masses $<10^{12.2}\,\rm M_{\odot}$;
(ii) at $M_{\rm stars}= 10^{10.8}-10^{11.1}\,\rm M_{\odot}$
centrals hosted by halos of masses $10^{12.7}\,\rm M_{\odot}-10^{14}\,\rm M_{\odot}$ 
are $\approx 35$\% more likely to be slow rotators than galaxies 
hosted by halos of masses $10^{12.2}\,\rm M_{\odot}-10^{12.7}\,\rm M_{\odot}$.
In Fig.~\ref{FSRMass4} we show BCGs separately because  
their properties (i.e. overly massive and star-forming) lead to most of them being fast rotators.

In the bottom panel of Fig.~\ref{FSRMass4} we show the effect of halo mass 
on the population of satellite galaxies.
We see no evident effect of environment. However, 
when studying the subsample of passive satellite galaxies (i.e. those 
with $\delta\,\rm MS <0.1$; see Eq.~\ref{MSdef} for a definition) we see a strong environmental effect. 
This is shown in Fig.~\ref{FSRMass4b}. 
We find that among passive satellites, $\rm F_{\rm SR}$ {\it increases 
with decreasing halo mass} at $M_{\rm stars}\lesssim 10^{11}\,\rm M_{\odot}$.
{The latter is clearly visible when we compare satellites in halos of masses below and above 
$10^{13}\,\rm M_{\odot}$.} 
At higher masses the statistics are too poor to draw any conclusion.
At first glance this result is unexpected, as the overall trend of satellites plus centrals shows 
more slow rotators in denser environments. 
We interpret this trend as due to passive satellites in low density environments being quenched 
at the same time as they go through a morphological transformation \citep{Trayford15,Dubois16}. The satellite population 
we are studying here are relatively massive galaxies, $M_{\rm stars}>10^{10}\,\rm M_{\odot}$, which 
are unlikely to be quenched solely by environment in halos of masses $<10^{13}\,\rm M_{\odot}$.
In more massive halos, $M_{\rm halo}\gtrsim 10^{14}\,\rm M_{\odot}$, 
galaxies can quench without morphological transformation, through e.g. ram pressure and/or tidal stripping. 
Thus, our simulations prediction that a trend with halo mass should be seen for satellite galaxies, but only 
in the subsample of passive satellites.
Selecting passive centrals increases the overall $\rm F_{\rm SR}$, but does not significantly change the 
halo mass effect we described above (not shown here).

\citet{Veale17} and \citet{Brough17} recently concluded that 
the dependence of $\rm F_{\rm SR}$ on environment is fully accounted for by the stellar mass of galaxies: 
more massive galaxies live in denser environments, and thus no environmental effects are seen at fixed stellar mass.
\citet{Brough17} focused exclusively on cluster environments, and thus their galaxy population was vastly dominated by 
satellite galaxies. As we showed here, \eagle\ and \Hydrangea\ show that environmental effects (as manifested 
through a halo mass dependence) in the satellite 
galaxy population are minimal, and even less obvious in the cluster population alone (see dashed and dot-dashed lines in 
the bottom panel of Fig.~\ref{FSRMass4}). 
On the other hand, we predict that the halo mass effect on slow rotators should be detectable 
in the population of centrals and passive satellite galaxies, but only if a wide range of halo masses is explored, 
$10^{11}\,\rm M_{\odot})\lesssim M_{\rm halo}\lesssim 10^{15}\,\rm M_{\odot}$.

\section{The physical origin of slow rotators}\label{physicalorigin}

\begin{figure}
\begin{center}
\includegraphics[trim=1mm 5mm 0mm 0mm, clip,width=0.48\textwidth]{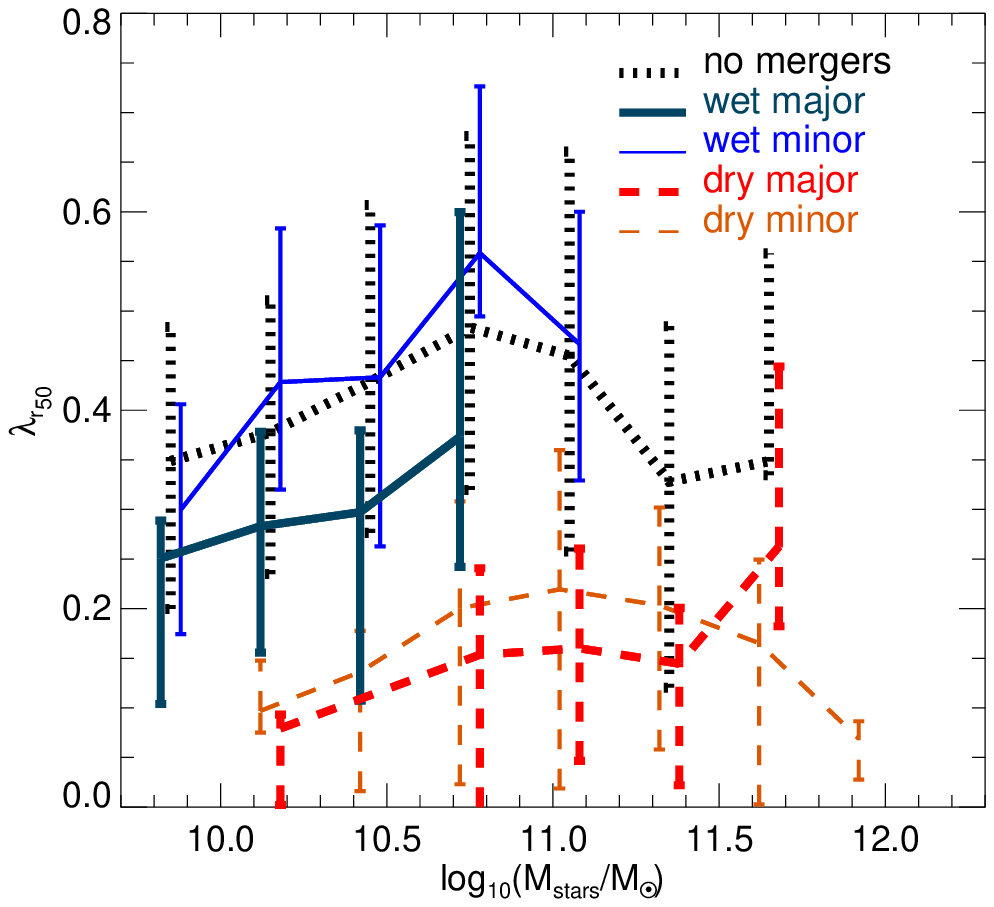}
\includegraphics[trim=1mm 5mm 0mm 0mm, clip,width=0.48\textwidth]{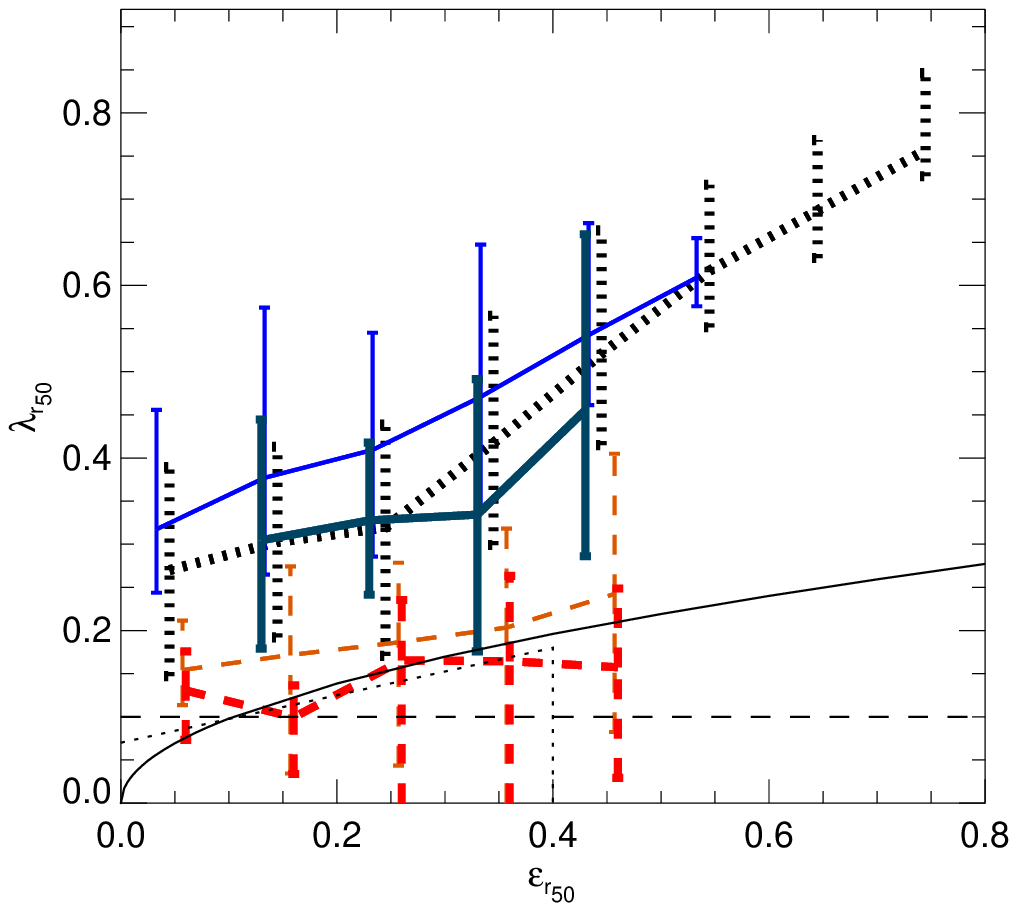}
\caption{$\lambda_{\rm r_{50}}$ as a function stellar mass (top panel) and 
$\epsilon_{\rm r_{50}}$ (bottom panel) for galaxies with $M_{\rm stars}>5\times 10^{9}\,\rm M_{\odot}$ 
in the Ref-L100N1504 simulation at $z=0$.
Lines with error bars show the medians and 
$25^{\rm th}-75^{\rm th}$ percentiles, respectively, for different samples of galaxies with different merger histories. 
{The latter is shown only for bins with $\ge 5$ galaxies.}
The samples correspond to galaxies that have not experienced mergers (thick dotted line), 
and that experienced at least one minor wet (thin solid line), minor dry (thin dashed line), 
major wet (thick solid line) and major dry (thick dashed line) merger in the last $10$~Gyr. In the case of minor 
mergers we selected galaxies that did not have any major mergers in the last $10$~Gyr. 
Here the separation between wet and dry merger is at $R_{\rm gas,merger}=0.1$ (see Eq.~\ref{fgasmerg} for a definition).
There is a clear 
connection between dry mergers (either major or minor) with slow rotation kinematics at 
$M_{\rm stellar}>10^{10}\,\rm M_{\odot}$.}
\label{SRMergers}
\end{center}
\end{figure}

\begin{figure}
\begin{center}
\includegraphics[trim=1mm 5mm 0mm 0mm, clip,width=0.495\textwidth]{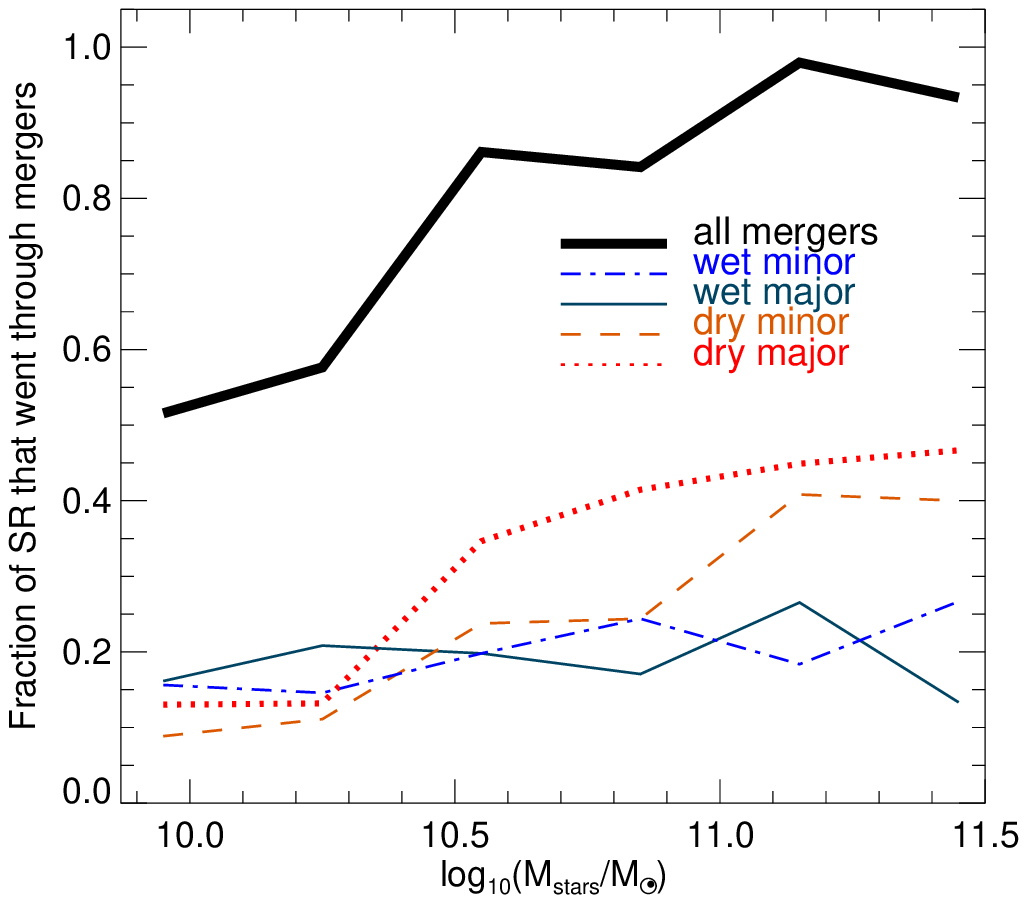}
\includegraphics[trim=1mm 5mm 0mm 0mm, clip,width=0.495\textwidth]{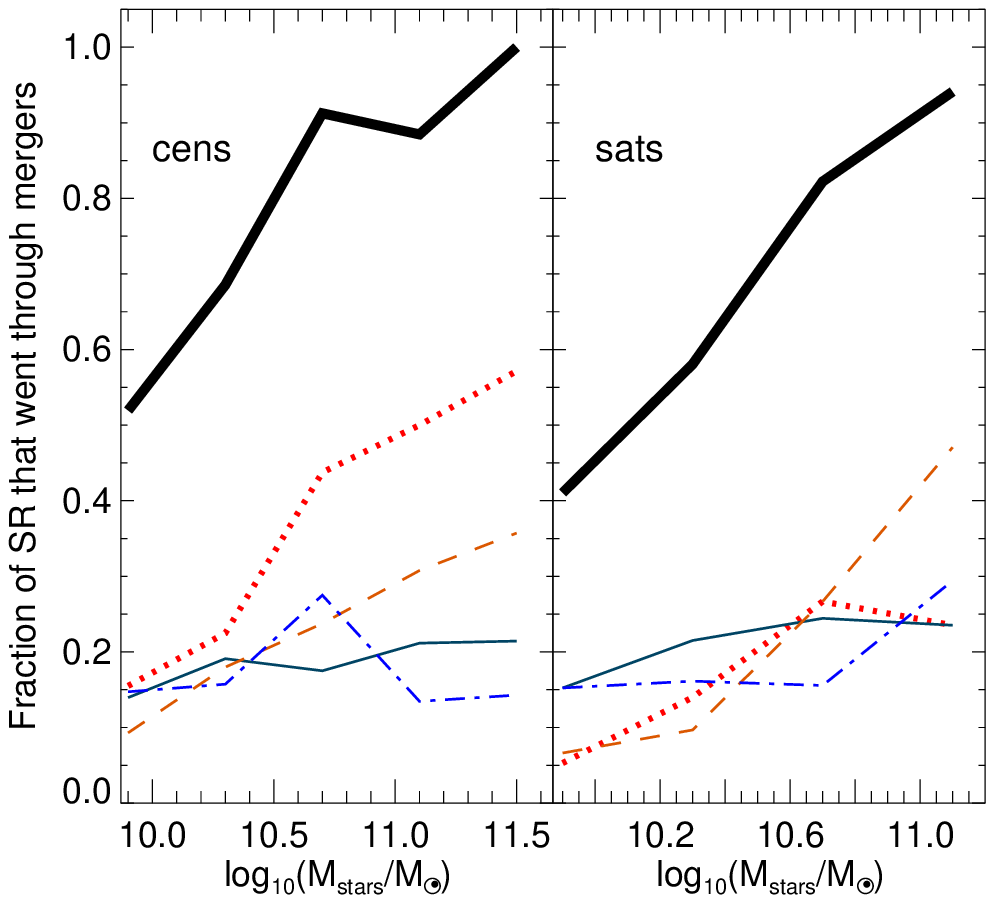}
\caption{{\it Top panel:} Fraction of slow rotators at $z=0$ in the Ref-L100N1504 simulation that suffered any 
type of mergers (thick solid line), and that have had at least one major dry (dotted line), major wet (thin solid line), 
minor dry (dashed line) or minor wet mergers (dot-dashed line), over the last $10$~Gyr, as a function of their $z=0$ stellar mass.
{We show only those bins with $\ge 10$ objects.}
We adopt the slow rotator classification of \citet{Cappellari16}. 
{\it Bottom panel:} 
As in the top panel but split into central (left) and satellite (right) galaxies. 
Note that the x-axis in the panel of satellites spans a smaller dynamic range. This is because 
there are very few satellite galaxies with masses above $10^{11}\,\rm M_{\odot}$.}
\label{SRMergersHistory}
\end{center}
\end{figure}

In this section we analyse the physical origin of slow rotators in two ways. 
First, we analyse the merger history of galaxies that at $z=0$ are slow rotators to 
establish how correlated their low $\lambda_{\rm R}$ is with 
the presence and type of mergers they suffered throughout their lives, if any. 
Second, we analyse the effect that individual merging events have on $\lambda_{\rm R}$ 
and $\epsilon$ by comparing the kinematic properties of the main progenitors  
and merger remnants. 
\citet{Lagos17} analysed the merger history of galaxies 
in the Ref-L100N1504 simulation, adding information on 
the cold gas masses of merging galaxies, orientation of mergers, orbital 
angular momentum and mass ratios. We use this extended merger catalogue  
to study the connection between mergers and slow rotation. Thus, here we 
focus solely on the Ref-L100N1504 simulation. 
We classify mergers as dry ($R_{\rm gas,merger}\le 0.1$), wet 
($R_{\rm gas,merger}> 0.1$ ), major (secondary to primary stellar mass ratio, $m_{\rm s}/m_{\rm p}\ge 0.3$) 
and minor ($0.1\le m_{\rm s}/m_{\rm p}< 0.3$; see $\S$~\ref{galmergerssec}).

We take all galaxies at $z=0$ in the Ref-L100N1504 simulation and split them into $5$ samples: (i) galaxies 
that have not suffered mergers, those that have not suffered major mergers, but have suffered 
either (ii) dry or (iii) wet minor mergers, and those that have had major mergers 
either (iv) dry or (v) wet. Galaxies that suffered major mergers could have also suffered minor mergers, 
but from the samples of minor mergers we remove all galaxies that had at least $1$ major merger. 
This is done under the premise that major mergers have a more important effect 
on galaxy properties than minor mergers. This is supported by our previous results (\citealt{Lagos17}).
Our selection is based on the merger history of galaxies over the last $10$~Gyr (i.e. approximately 
since $z=2$).

Fig.~\ref{SRMergers} shows the median $\lambda_{\rm r_{\rm 50}}$ as a function of $M_{\rm stars}$ 
and $\epsilon_{\rm r_{\rm 50}}$ for galaxies at $z=0$ that have $M_{\rm stars}>5\times 10^9\,\rm M_{\odot}$, 
separated in the $5$ samples above, i.e. depending on their merger history. 
Here we do not distinguish between recent or far in the past mergers, but simply 
count their occurrence. We see a clear connection between the incidence of dry mergers 
(either major or minor) with slow rotation in galaxies 
with $M_{\rm stars}> 10^{10}\rm\,M_{\odot}$. On average, 
galaxies that went through dry major mergers have a lower $\lambda_{\rm r_{\rm 50}}$ 
than those that went through dry minor mergers. 
The remnants of wet major mergers 
also tend to have relatively low $\lambda_{\rm r_{\rm 50}}$, but not enough to place them 
on the slow rotation class, though $\approx 10$\% of the wet major merger sample are slow rotators.
 Galaxies that had wet minor mergers have slightly larger $\lambda_{\rm r_{\rm 50}}$ at fixed $\epsilon_{\rm r_{\rm 50}}$ 
than galaxies that have not had mergers, possibly reflecting the fact that the former are on 
average a lot more gas rich (average neutral 
gas to stellar mass ratio of $19$\% compared to $6$\% in the latter sample). 
In \citet{Lagos16b} we showed, also using \eagle, that continuous gas accretion and star formation 
efficiently spin up galaxies because the angular momentum brought by newly accreted gas is expected to grow proportionally 
with time \citep{Catelan96a}. 
Regardless of this effect, we find that the parameter space of $\lambda_{\rm r_{\rm 50}}\gtrsim 0.7$ and $\epsilon \gtrsim 0.6$ 
is almost exclusively occupied by galaxies that have not had any mergers.  

The fact that wet minor mergers appear to only slightly affect galaxies 
agrees with the conclusions of \citet{Lagos17}, in which it was shown 
that galaxies undergoing wet minor mergers have angular momentum radial profiles  
 similar to galaxies that have not had mergers. The exception is the very centres 
of those galaxies, as the remnants of wet minor mergers tend to have slightly more 
massive bulges (see their Fig. 6).
Although there is a clear trend between how galaxies populate 
the $\lambda_{\rm r_{\rm 50}}-M_{\rm stars}$
and $\lambda_{\rm r_{\rm 50}}-\epsilon_{\rm r_{\rm 50}}$ planes and their merger history, 
the scatter is large, suggesting that mergers result in a plethora of remnants 
with no unique outcome. Our results support the findings of \citet{Naab14} though 
with $\approx 50$ times more mergers, which allows us to disentangle 
preferred formation mechanisms.

To disentangle the formation paths of slow rotators in \eagle, we focus on their merger history 
as a function of stellar mass.
The top panel of Fig.~\ref{SRMergersHistory} shows the fraction of slow rotators that 
went through the $4$ merging scenarios described above (wet/dry minor mergers, 
wet/dry major mergers), as a function of stellar mass at $z=0$. 
We also show as black lines the fraction of slow rotators that had any form of merger 
with $m_{\rm s}/m_{\rm p}\ge 0.1$.
We define slow rotators using the \citet{Cappellari16} criterion.
 
At $10^{10}\,\rm M_{\odot}\lesssim M_{\rm stars}\lesssim 3\times 10^{10}\,\rm M_{\odot}$, {\it $\approx 40$\% of slow rotators have not had 
any mergers}. This percentage decreases systematically with increasing stellar mass, and 
by $M_{\rm stars}>10^{11}\,\rm M_{\odot}$, $96$\% of the slow rotators had at least one merger 
during their past $10$~Gyr. Among the slow rotators that had mergers, the most common type of merger is 
dry major merger, followed by minor mergers and wet major mergers.

\begin{figure}
\begin{center}
\includegraphics[trim=1mm 7mm 0mm -5.5mm, clip,width=0.495\textwidth]{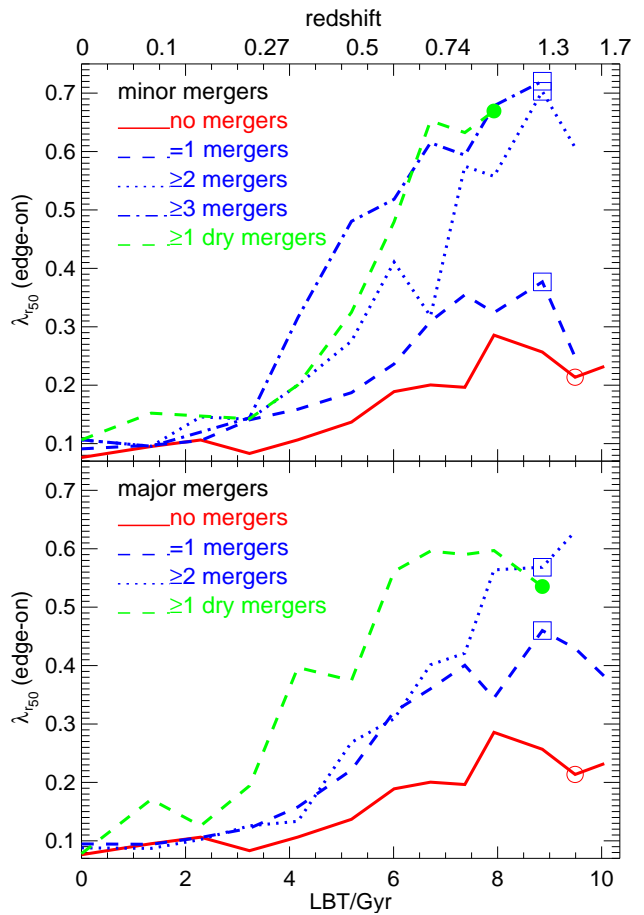}
\caption{{\it Top panel:} Median $\lambda_{\rm r_{50}}$ history of galaxies classified as slow rotators at $z=0$ 
that have $M_{\rm stars}\ge 10^{10}\,\rm M_{\odot}$ in the Ref-L100N1504 simulation and that experienced only 
minor or no mergers.
Here $\lambda_{\rm r_{50}}$ was measured orienting galaxies edge-on. 
We show the history of these galaxies split into $5$ samples: 
(i) slow rotators that have not experienced a merger (solid line), 
(ii) slow rotators that experienced one merger (blue dashed line), 
(iii) at least two mergers (blue dotted line),  
(iv) at least 3 mergers (blue dot-dashed line),
and (v) at least one dry merger (green dashed line), as labelled. In the case of samples 
(ii), (iii) and (iv) we do not distinguish by the gas fraction of the merger, 
while in the case of (v) we impose a merger gas to stellar mass ratio 
threshold of $0.1$ (see Eq.~\ref{fgasmerg} for a definition). 
For reference, symbols show the median mass-weighted stellar age of the galaxies in each sample. 
{\it Bottom panel:} As in the top panel but for slow rotators that experienced major mergers. 
In this case the progenitors of these slow rotators could have also experienced minor mergers. 
Here we show samples (i), (ii), (iii) and (v) because major mergers are less frequent and thus, the sample 
of galaxies with $\ge 3$ major mergers is too small {($6$ galaxies)}.}
\label{SRMergersProgenitors}
\end{center}
\end{figure}

In the bottom panel of Fig.~\ref{SRMergersHistory} we separate centrals and satellites. 
The prevalence of dry major mergers is more significant in central galaxies. 
Here, dry major mergers are twice more common in slow rotators than the other forms of mergers.
For satellites we see that the different types of mergers have a similar incidence 
and dry minor mergers become more prevalent at $M_{\rm stars}\gtrsim 10^{10.6}\,\rm M_{\odot}$.
This shows that the importance of mergers and their type for slow rotation may have an environmental 
dependence.
Nevertheless, there is a clear connection between dry mergers and 
slow rotators, but we still need to establish whether there is a causal connection between the two. 
We come back to this in $\S$~\ref{MergersVsSlowR} where we 
analyse the effect of individual merger events on 
 $\lambda_{\rm R}$ and $\epsilon_{\rm r_{\rm 50}}$.

In Fig.~\ref{SRMergersProgenitors} we show the history of 
$\lambda_{\rm r_{50}}$ of galaxies that at $z=0$ 
are classified as slow rotators. For the latter we apply 
a simple cut of $\lambda_{\rm r_{50}} \le 0.1$ \citep{Emsellem07}.
To make the interpretation easier, we show the history of 
$\lambda_{\rm r_{50}}$ measured after orienting galaxies edge-on (i.e. $\lambda_{\rm r_{50}}$ 
takes its maximum value).
We separate slow rotators that have only had minor mergers (top panel),
and that have had major mergers (bottom panel). 
The latter could also have had minor mergers. 
In addition to minor and major mergers, we distinguish between different 
numbers of mergers (either wet or dry), and also show separately the slow rotators that had dry mergers.
Symbols show the median mass-weighted stellar age of the galaxies in the different samples.
Slow rotators that have not experienced any minor or major mergers were born with low 
$\lambda_{\rm r_{50}}$ values, and at a look-back time of $8.5$~Gyr, which is roughly 
the median mass-weighted stellar age of all these galaxies, 
they have $\lambda_{\rm r_{50}}$ at least twice smaller than the rest of the galaxies.
This is driven by the environments in which these galaxies formed. We come back to this in 
$\S$~\ref{spinparameterDM}.

The top panel of Fig.~\ref{SRMergersProgenitors} shows that there is a cumulative effect of minor mergers, 
in which galaxies could have started with a high $\lambda_{\rm r_{50}}$ but lost it through successive minor 
merger events. Note that those slow rotators that only had one minor merger, started with relatively low 
$\lambda_{\rm r_{50}}$. The subsample of slow rotators that had at least one dry minor merger shows
the most dramatic evolution of $\lambda_{\rm r_{50}}$ {(i.e. the fastest decrease)}, again supporting our conclusion 
that dry mergers are most effective at producing slow rotators.
{In the case of galaxies having had $\ge 3$ minor mergers, a fast decrease of 
$\lambda_{\rm r_{50}}$ is also seen, but this sample includes only $13$ galaxies at $z=0$.}
\citet{Penoyre17} recently analysed the Illustris simulation 
and concluded that they do not see a cumulative effect of minor mergers, in contradiction 
with the findings of \citet{Naab14} and our results here.  
Given how sensitive the outcome of mergers are to their gas fraction (see $\S$~\ref{MergersVsSlowR}), 
one possible explanation to the different findings is that \eagle\ produces a different 
gas fraction evolution of galaxies compared to Illustris, impacting the effect mergers have on galaxies. 
However, because the nature of these simulations is 
complex, with many processes acting simultaneously at any one time, it is hard to conclusively say what drives the differences 
between \eagle\ and Illustris.

The bottom panel of Fig.~\ref{SRMergersProgenitors} shows that {single} major mergers generally have a stronger 
effect than {single} minor mergers on the history of $\lambda_{\rm r_{50}}$. This is clear when comparing the 
dashed lines between the top and bottom panels of Fig.~\ref{SRMergersProgenitors}, 
where galaxies that went through one major merger started with 
$\lambda_{\rm r_{50}}\approx 0.45$, on average, while those that went through one minor merger 
started with $\lambda_{\rm r_{50}}\approx 0.35$, on average.
Major mergers also display a cumulative effect, but given how much rarer they are compared 
to minor mergers (see Fig.~$2$ in \citealt{Lagos17}), the significance of this is minimal for the entire 
galaxy population; i.e. there are only $11$ galaxies in the entire simulated 
volume that had $\ge 2$ major mergers in the last $10$~Gyr. When selecting 
slow rotators that had at least one dry major merger, we see a much more drastic decrease 
in $\lambda_{\rm r_{50}}$. 
In $\S$~\ref{MergersVsSlowR} we show that dry mergers are connected with the 
most significant decrease in $\lambda_{\rm r_{50}}$ in individual merger events.

For both minor and major mergers, we see that slow rotators that went through dry mergers, experience a rapid decrease of $\lambda_{\rm r_{50}}$ 
at look-back times $\lesssim 6$~Gyr. This is due to the dry merger rate increasing rapidly  
after that epoch towards $z=0$. On the other hand, the total merger rate decreases smoothly, which explains why 
the $\lambda_{\rm r_{50}}$ evolutionary tracks of galaxies that suffered one or two mergers display a smoother decrease.

\begin{figure}
\begin{center}
\includegraphics[trim=1mm 15.5mm 0mm -4.7mm, clip,width=0.495\textwidth]{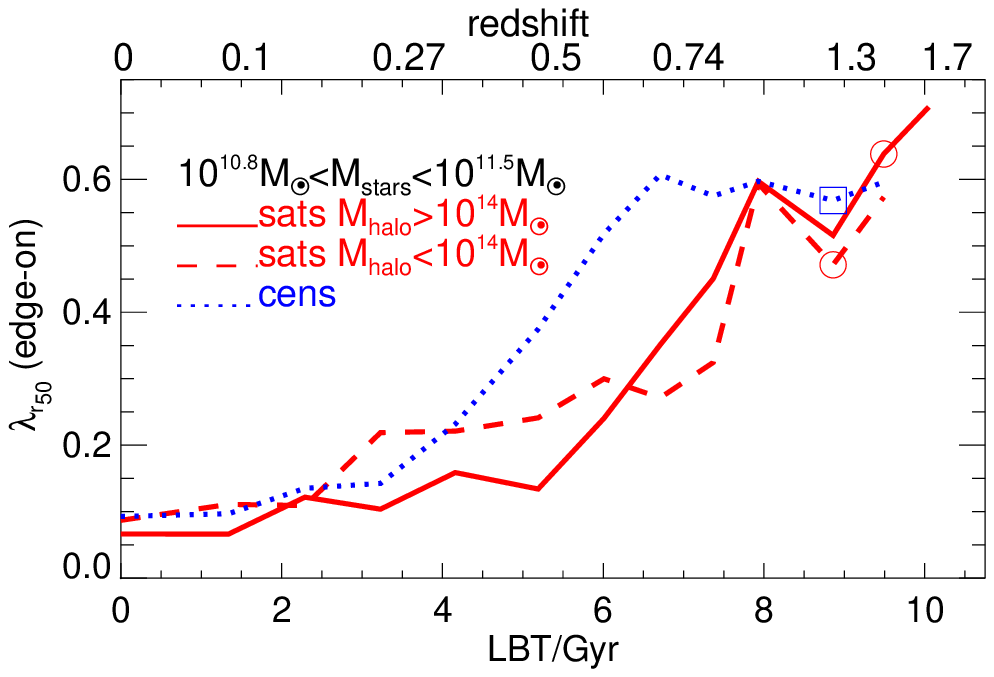}
\includegraphics[trim=1mm 7mm 0mm 2.5mm, clip,width=0.495\textwidth]{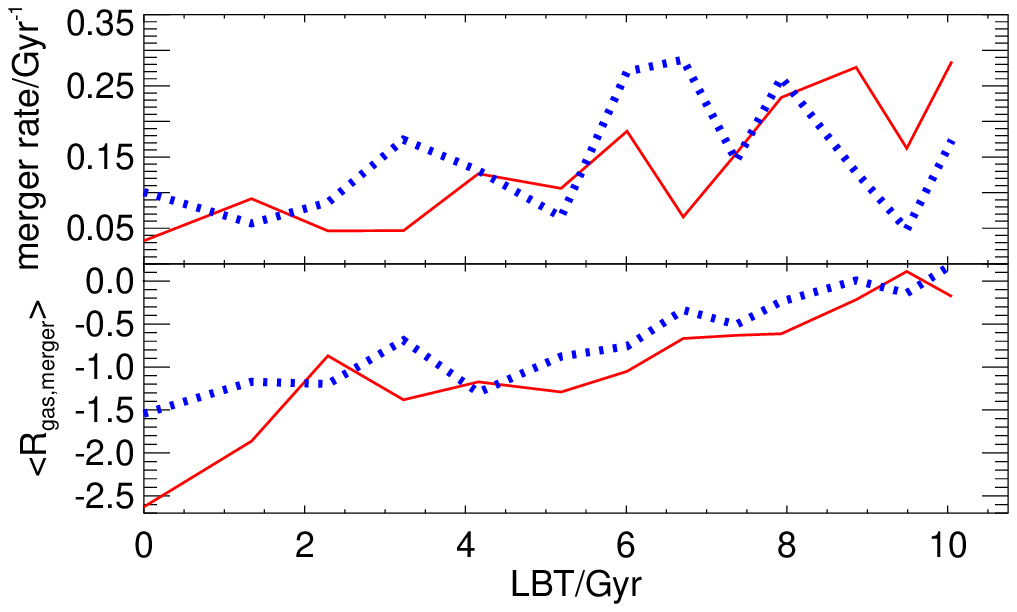}
\caption{{ {\it Top panel:} Median $\lambda_{\rm r_{50}}$ history of galaxies that at $z=0$ have 
$\lambda_{\rm r_{\rm 50}}\le 0.1$ and $10^{10.8}\rm\, M_{\odot}<M_{\rm stars}<10^{11.5}\rm\, M_{\odot}$ in the Ref-L100N1504 simulation.
We separate galaxies in this mass bin into satellites that at $z=0$ are hosted by halos of masses above (solid line) and below (dashed line) $10^{14}\rm\, M_{\odot}$, 
and centrals (dotted line). 
Symbols show the median mass-weighted stellar age of the galaxies in each sample. 
{\it Middle panel:} merger rate of the galaxies in the top panel, separating into 
centrals and satellites. {\it Bottom panel:} median neutral gas-to-stellar mass ratio 
of the mergers in the middle panel.}}
\label{SRMergersProgenitors2}
\end{center}
\end{figure}

The fact that all the galaxies that at $z=0$ are slow rotators display an overall spin down throughout 
their lives even in the absence of mergers, is probably connected to the evolution of the local environment 
in which galaxies and halos reside. \citet{Welker15} show that halos, as they move from 
 high-vorticity regions in the cosmic web towards the filaments and nodes, start to be subject to less 
and less coherent gas accretion. In the limit of nodes in the cosmic web, accretion happens a lot more isotropically 
than in the high-vorticity regions or filaments, 
with several filaments connecting to the node from different directions. High-vorticity regions accrete
gas from preferential directions, thus gaining more 
coherent angular momentum. The overall spin down we see in massive galaxies and halos (see $\S$~\ref{spinparameterDM}) is most likely 
linked to the overall environmental evolution. Fast rotators do not necessarily experience the same spin down because they tend to inhabit 
lower mass halos, which are less clustered.

{ In $\S$~\ref{VSselec} we showed that satellites are $25$\% more likely to be slow rotators than centrals 
at $M_{\rm stars}\gtrsim 10^{10.8}\rm\, M_{\odot}$. Fig.~\ref{SRMergersHistory} shows that at $10^{10.8}\rm\, M_{\odot}<M_{\rm stars}<10^{11.5}\rm\, M_{\odot}$, satellite 
galaxies have a slightly higher merger incidence than centrals ($92\%$ vs. $87$\%). 
The top panel of Fig.~\ref{SRMergersProgenitors2} shows that satellites that at $z=0$ are slow rotators, spun down 
earlier (at look-back times $<8$~Gyr) and have older stellar populations than centrals (which spun down at $<6$~Gyr), at fixed stellar mass. 
The latter becomes exacerbated in satellites of halos with masses $>10^{14}\,\rm M_{\odot}$. 
The middle panel of Fig.~\ref{SRMergersProgenitors2} shows that this earlier spinning down is due to the satellite merger 
rate peaking at higher redshifts than centrals. 
In addition, the galaxy mergers suffered by the population of $z=0$ satellite 
slow rotators were more gas poor than those suffered by centrals, on average (bottom panel of Fig.~\ref{SRMergersProgenitors2}). 
As the merger gas fraction is correlated with the resulting change in $\lambda_{\rm R}$ (which we show in $\S$~\ref{MergersVsSlowR}),  
it is expected that the satellite mergers have a more devastating effect on $\lambda_{\rm R}$, on average, than the mergers centrals experience. 
The difference in $R_{\rm gas,merger}$ between centrals and satellites holds when we analyse 
the overall population at $10^{10.8}\rm\, M_{\odot}<M_{\rm stars}<10^{11.5}\rm\, M_{\odot}$ (i.e. regardless of their $z=0$ $\lambda_{\rm R}$). 
Thus, a higher $F_{\rm SR}$ in satellites at $z=0$ in \eagle\ can be connected to them having suffered slightly more mergers, and that were on average more 
gas poor than those centrals experienced.}

\subsection{The effect of individual merger events on $\lambda_{\rm R}$}\label{MergersVsSlowR}

\begin{figure}
\begin{center}
\includegraphics[trim=3mm 6.5mm 0mm 13mm, clip,width=0.495\textwidth]{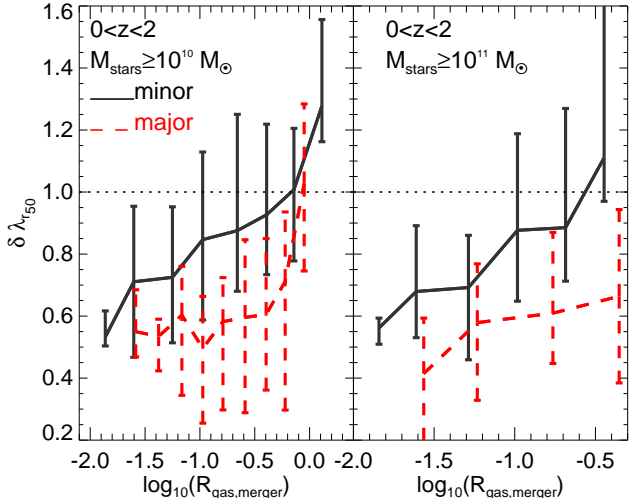}
\caption{Variation of $\lambda_{\rm r}$ measured at $r_{\rm 50}$ 
as a function of the gas to stellar mass ratio of the merger (see Eq.~\ref{fgasmerg}) 
separating minor and major mergers, as labelled. The left panel shows all the
mergers that took place in galaxies with $M_{\rm stars}\ge 10^{10}\,\rm M_{\odot}$ 
at $0\le z\le 2$, while the right panel shows the subsample 
of galaxies with $M_{\rm stars}\ge 10^{11}\,\rm M_{\odot}$. 
Lines with error bars show the median and $25^{\rm th}-75^{\rm th}$ 
percentile ranges. Only bins with $\ge 5$ objects are shown.
For reference, the dotted horizontal line shows no change in 
$\lambda_{\rm r}$. Positive values indicate the merger remnant has a higher 
value of $\lambda_{\rm r}$ than the progenitor.}
\label{LambdaRvsMergers}
\end{center}
\end{figure}

\begin{figure}
\begin{center}
\includegraphics[trim=3mm 6.5mm 0mm 13mm, clip,width=0.495\textwidth]{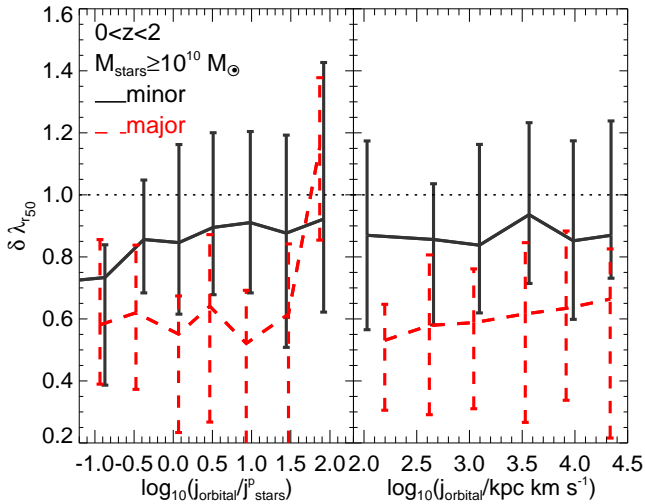}
\caption{{ $\delta\,\lambda_{\rm R}$ as a function of the ratio between the orbital 
and stellar specific angular momentum of the primary galaxy (left panel), 
and the orbital specific angular momentum (right panel), for all the mergers 
that took place in galaxies with $M_{\rm stars}\ge 10^{10}\,\rm M_{\odot}$ 
at $0\le z\le 2$. We separate minor and major mergers, as labelled.}}
\label{LambdaRvsMergers2}
\end{center}
\end{figure}

In order to determine the effect that individual mergers have on 
the rotation of galaxies, we take all the minor and major mergers 
that have primary galaxies with $M_{\rm stars} \ge 10^{10}\,\rm M_{\odot}$ 
from $z=0$ to $z=2$, and compute the change in $\lambda_{\rm R}$ before and after the merger. 
Note that here we do not distinguish between descendants that are slow/fast rotators, 
but take all mergers.
We then compare $\lambda_{\rm R}$ between the main progenitor (the most massive) in the last
snapshot the two merging galaxies were identified individually 
and the merger remnant. The latter corresponds to the first snapshot in which the two galaxies appear
merged. Typically the timescale between these snapshots is $\approx 0.5$~Gyr.
We define 

\begin{equation}
\delta\,\lambda_{\rm R}=\frac{\lambda_{\rm R,rem}}{\lambda_{\rm R,prog}}, 
\end{equation}

\noindent with $\lambda_{\rm R,rem}$ and $\lambda_{\rm R,prog}$ being the remnant's and main progenitor's 
$\lambda_{\rm R}$, respectively. 

Fig.~\ref{LambdaRvsMergers} shows $\delta\,\lambda_{\rm r_{\rm 50}}$ (measured at $r_{50}$) 
as a function of the cold gas to stellar mass ratio of the merger, $R_{\rm gas,merger}$ 
(Eq.~\ref{fgasmerg}). We show minor and major mergers separately.  
The right panel of Fig.~\ref{LambdaRvsMergers} 
shows the subsample of galaxies with $M_{\rm stars} \ge 10^{11}\,\rm M_{\odot}$.
There is a positive correlation between 
$\delta\,\lambda_{\rm r_{\rm 50}}$ and $R_{\rm gas,merger}$, 
{but with an offset in normalization in a way that major mergers 
are $25$\% more likely to decrease $\lambda_{\rm R}$ compared to minor mergers.
Major mergers also do this to a greater extent than minor mergers, 
decreasing $\lambda_{\rm R}$ by $39$\% compared to $12$\% in the latter, on average.}
Galaxy minor {(major)} mergers with $R_{\rm gas,merger}\gtrsim 0.5$ {($0.8$)} have a clear preference for 
increasing $\lambda_{\rm R}$, while those with $R_{\rm gas,merger}\lesssim 0.1$ 
have a {strong} preference for decreasing $\lambda_{\rm R}$. However, 
the scatter around the median relation is large, suggesting that the 
effect of a merger on $\lambda_{\rm R}$ is not uniquely determined by its mass ratio and 
gas fraction. 
\citet{Penoyre17} found in Illustris that major mergers, regardless of their gas fraction, are connected  
with the spinning down of galaxies, in contradiction with our findings. This may be due 
to their major mergers being mostly gas poor (gas-to-stellar mass ratios $\lesssim 0.1$; see their Fig. $13$),
thus, lacking the very gas-rich major mergers we obtain in \eagle\ that spin up galaxies ($R_{\rm gas,merger} \gtrsim 0.8$).

Focusing specifically on dry mergers ($R_{\rm gas,merger}\le 0.1$), 
we find that in {$\approx 15$\%} of the major mergers $\lambda_{\rm R}$ increases,
while for minor mergers this fraction is {$25$\%}. 
Selecting only massive galaxies in \eagle\ 
(right panel in Fig.~\ref{LambdaRvsMergers})
does not change the correlation between 
$\delta\,\lambda_{\rm r_{\rm 50}}$ and $R_{\rm gas,merger}$ significantly.
We analysed $\delta\,\lambda_{\rm 2r_{\rm 50}}$ (measured at $2\,r_{\rm 50}$) 
and found a very similar relation to that shown in Fig.~\ref{LambdaRvsMergers}. This suggests 
that mergers modify $\lambda_{\rm R}$ in a similar fashion over a large radial range.
It is clear that the high incidence of dry major mergers 
in the slow rotator population of Fig.~\ref{SRMergersHistory} 
 is due to these mergers having a detrimental effect on 
$\lambda_{\rm R}$, on average.

{Fig.~\ref{LambdaRvsMergers2} shows $\delta\,\lambda_{\rm R}$ 
as a function of $j_{\rm orbital}/j^{\rm p}_{\rm stars}$ (left panel)
 and $j_{\rm orbital}$ (right panel). Here, $j^{\rm p}_{\rm stars}$ is the total stellar 
specific angular momentum of the primary galaxy.
In major mergers the orbital angular momentum has an effect on $\lambda_{\rm R}$, which is more clearly seen 
when we study ${j}_{\rm orbital}/j^{\rm p}_{\rm stars}$, in a way that 
high ${j}_{\rm orbital}$ drives smaller changes in $\lambda_{\rm R}$. Minor mergers display a much weaker 
dependence on $j_{\rm orbital}/j^{\rm p}_{\rm stars}$ and no clear dependence on $j_{\rm orbital}$. 
We also studied the effect of alignments of the rotation axis of the merger pair 
and found no effect on $\lambda_{\rm R}$ (not shown here).

\citet{Li18} analysed the effect of the merger orbits on the shape and $\lambda_{\rm R}$ of merger remnants using the 
Illustris simulation, and found that circular orbits tend to produce fast rotators, while radial orbits 
 produce slow rotators. In our calculation, radial orbits correspond to low $j_{\rm orbital}$, 
and in agreement with Li et al., we find that the decrease in $\lambda_{\rm R}$ is the largest in these cases. 
However, the scatter around the median is very large, and the dependence on $R_{\rm gas,merger}$ 
is stronger. This agrees with the conclusion of \citet{Lagos17}, who showed that the gas fraction of the merger 
is the most fundamental property determining the effect on the angular momentum of the merger remnant in \eagle, with the 
mass ratio modulating the strength of the effect.}

We find that in the absence of mergers, galaxies display little change in their 
$\lambda_{\rm R}$, $<5$\%. 
This seems to contradict the result of \citet{Choi17}, who argued that 
most of the spin down of galaxies is driven by environment and not mergers.
This could be due to their study being performed exclusively on cluster regions, 
which represent an upper limit for the effect of environment. 

We also studied the effect of mergers on the ellipticity, $\epsilon$, 
of galaxies and found little effect (not shown here). 
Dry mergers have a tendency to increase $\epsilon$, which,  
combined with the fact that they tend to decrease 
$\lambda_{\rm R}$, results in galaxies ending up more comfortably in the slow rotator zone 
in the $\lambda_{\rm R}-\epsilon$ plane. On the other hand, wet major mergers 
tend to decrease $\epsilon$, thus making galaxies rounder. This is expected 
since wet mergers tend to increase the central stellar density of galaxies 
due to efficient gas fueling to the centre (e.g. \citealt{Cox06};\citealt{Robertson06}; 
\citealt{Johansson09}; \citealt{Peirani10}; \citealt{Moreno15}; \citealt{Lagos17}).

\subsection{The connection between slow rotators and the halo spin parameters}\label{spinparameterDM}

\begin{figure}
\begin{center}
\includegraphics[trim=1mm 4mm 0mm 3mm, clip,width=0.43\textwidth]{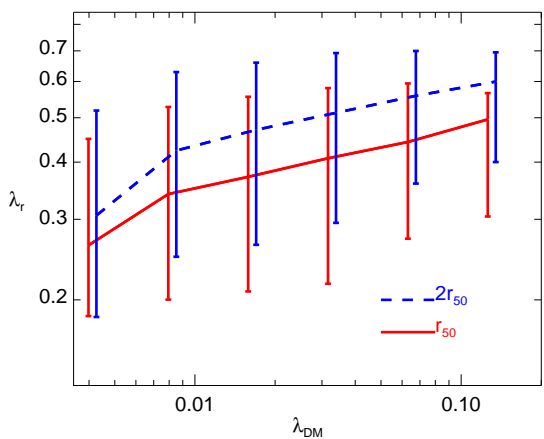}
\caption{{$\lambda_{\rm R}$ measured 
within $r_{\rm 50}$ and $2r_{50}$, as labelled, as a function of the 
dark matter halo $\lambda$ (Eq.~\ref{jhalo}) for central galaxies in \eagle\ at $z=0$
that have $M_{\rm stars}\ge 10^{10}\,\rm M_{\odot}$. Lines with error bars show the median 
and $16^{\rm th}-84^{\rm th}$ percentile ranges, respectively.}}
\label{LambdaHaloLambdaR}
\end{center}
\end{figure}

\begin{figure}
\begin{center}
\includegraphics[trim=1mm 15mm 0mm 0mm, clip,width=0.43\textwidth]{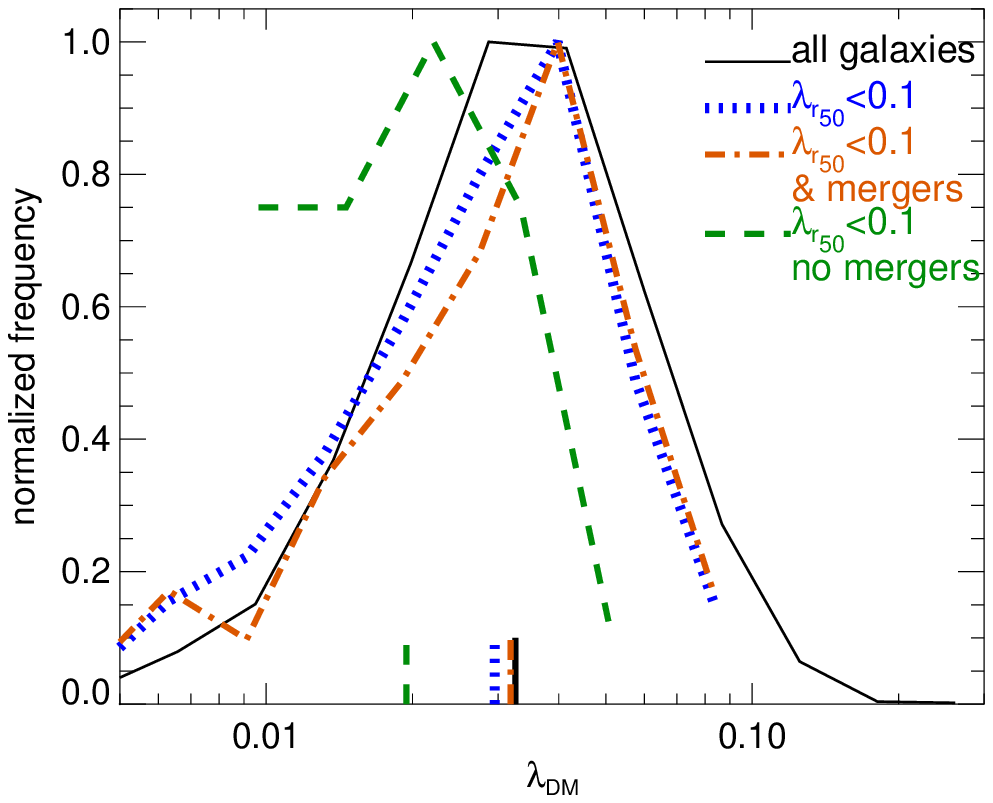}
\includegraphics[trim=1mm 6mm 0mm 2.3mm, clip,width=0.43\textwidth]{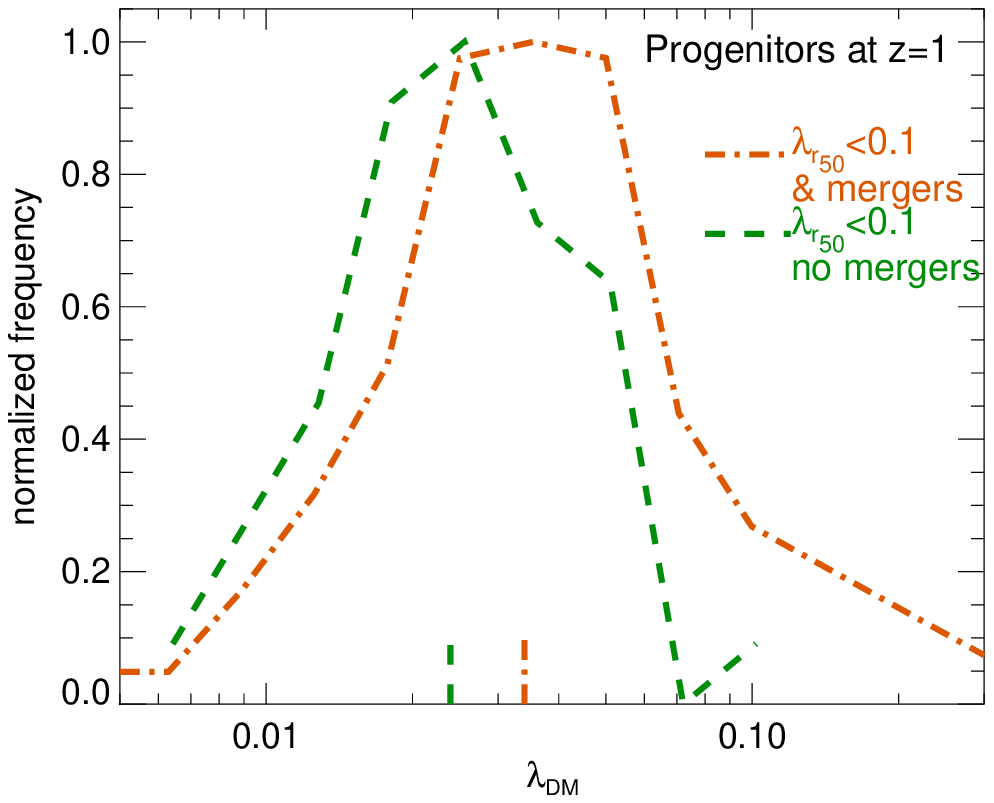}
\caption{{\it Top panel:} The dark matter halo $\lambda$ for central galaxies in \eagle\ at $z=0$
that have $M_{\rm stars}\ge 10^{10}\,\rm M_{\odot}$. Four subsamples are presented:
all galaxies (solid line), those with $\lambda_{\rm r_{50}}<0.1$ (dotted line), 
those with $\lambda_{\rm r_{50}}<0.1$ and that had at least 1 merger in the last $10$~Gyr 
 (dot-dashed line), and those with those with $\lambda_{\rm r_{50}}<0.1$ that have not 
had any mergere in the same period of time (dashed line). 
{\it Bottom panel:} As in the top panel but for the progenitor halos at $z=1$ of the two populations of
slow rotators at $z=0$: (i) those that experienced mergers, and (ii) those that did not.
We find that the host halos of slow rotators that at $z=0$ have not yet experienced mergers,
are biased towards low spins even at $z=1$.}
\label{LambdaHalo}
\end{center}
\end{figure}

\begin{figure}
\begin{center}
\includegraphics[trim=1mm 6mm 0mm 2mm, clip,width=0.49\textwidth]{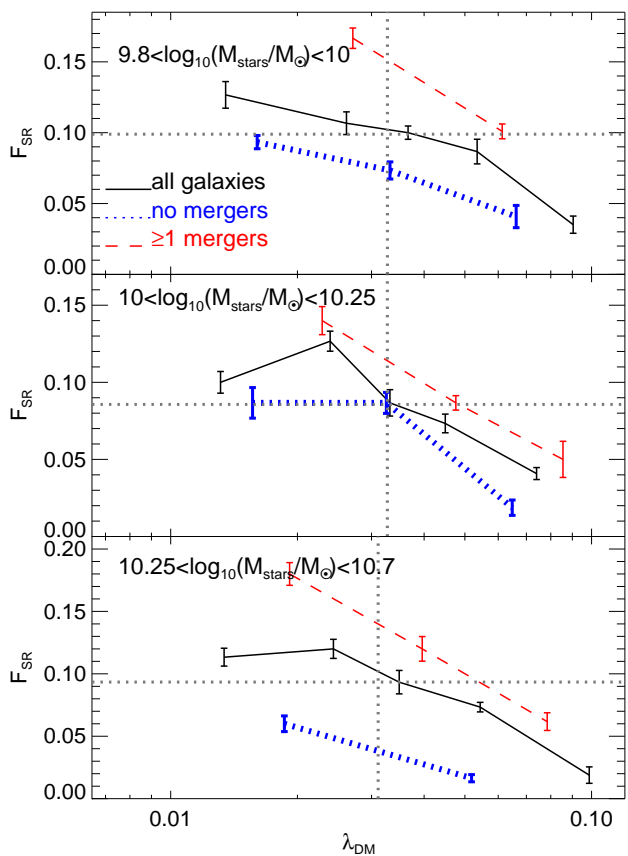}
\caption{The fraction of central galaxies that are slow rotators at $z=0$ (defined as those 
with $\lambda_{\rm r_{50}}\le 0.1$) 
as a function of the halo spin parameter, $\lambda_{\rm DM}$, in three bins of stellar mass 
of the central galaxy, as labelled in each panel. We show this for $3$ samples: all central galaxies (solid line), 
and the subsamples that had at least 
$1$ merger (dashed line) or no mergers (dotted line). Bins are chosen to have $\approx 150$ galaxies. 
Error bars correspond to $1$~standard deviation calculated 
with $10$ jackknife resamplings in individual mass bins. 
The horizontal and vertical line shows the fraction of slow rotators for all galaxies at $z=0$ in the stellar mass bins and their median 
$\lambda_{\rm DM}$.
There is a clear tendency for low $\lambda$ halos to have a higher fraction of slow 
rotators.}
\label{LambdaHaloFSR}
\end{center}
\end{figure}

Fig.~\ref{SRMergersHistory} showed that about $30$\% of the slow rotator population in the 
Ref-L100N1504 have not had any mergers. Fig.~\ref{SRMergersProgenitors} showed that 
these slow rotators also had modest $\lambda_{\rm r_{50}}$ in the past, smaller than the 
$\lambda_{\rm r_{50}}$ values of the progenitors of slow rotators that experienced mergers.
Here, we study the halos of these galaxies to understand why they are slow rotators.

We calculate the spin of halos, $\lambda_{\rm DM}$, as in \citet{Mo98},

\begin{equation}
\lambda_{\rm DM} = j_{\rm h}\, \frac{(10\,H)^{1/3}}{\sqrt{2}\,G^{2/3}}\,M^{-2/3}_{\rm h}
\label{jhalo}
\end{equation}

\noindent where $j_{\rm h}$ and $M_{\rm h}$ are the halo specific angular momentum and dark matter $M_{\rm 200}$ mass\footnote{Measured with 
all the dark matter particles within the halo's $r_{\rm 200}$, the radius within which the 
density is $200\,\rho_{\rm crit}$, with $\rho_{\rm crit}$ being the critical density.}, respectively, 
 $G$ is Newton's gravity constant and $H$ is the Hubble parameter. 
We calculate $j_{\rm h}$ with all the dark matter particles within a halo's $r_{\rm 200}$.
{We find a positive correlation between 
between the stellar $\lambda_{\rm R}$ and $\lambda_{\rm DM}$ in central galaxies (Fig.~\ref{LambdaHaloLambdaR}), 
but with significant scatter. Interestingly, this scatter tends to decrease with increasing 
aperture within which $\lambda_{\rm R}$ is measured.}

We now focus only on slow rotators to investigate the possible connection 
with their host halo spin.
The top panel of Fig.~\ref{LambdaHalo} shows the distribution of halo dark matter spin parameters, $\lambda_{\rm DM}$,  
of all central galaxies in the Ref-L100N1504 at $z=0$ that have stellar masses 
$>10^{10}\,\rm M_{\odot}$ (solid line). In the top panel of Fig.~\ref{LambdaHalo} 
we also show central galaxies with $\lambda_{\rm r_{50}}<0.1$ (dashed line), 
and the subsamples of these slow rotators that have had mergers (dot-dashed line) and had not had minor/major mergers (dashed line) 
over the last $10$~Gyr. 
Slow rotators that have not had mergers display a $\lambda_{\rm DM}$
 distribution that is significantly shifted compared to the other $3$ samples. Note that 
the median $\lambda_{\rm DM}$ of galaxies with $\lambda_{\rm r_{50}}<0.1$ that have 
had mergers is very similar to the overall population of central galaxies. The 
sample of centrals with $\lambda_{\rm r_{50}}<0.1$ has a slightly smaller median, but that 
is caused by the contribution of centrals with $\lambda_{\rm r_{50}}<0.1$ 
that have not had mergers. The latter is clear when 
comparing the slow rotators that have had mergers to the overall galaxy population (dot-dashed and solid lines in the top panel of Fig.~\ref{LambdaHalo}).
The median $\lambda_{\rm DM}$ of slow rotators that have not experienced mergers is 
a factor of $\approx 2$ smaller.
This explains why they formed with low $\lambda_{\rm r_{50}}$ values: they 
formed and evolved in halos of low spins. On average, galaxies and their host halos grow their angular momentum 
together in a way that resembles weak conservation of angular momentum \citep{Mo98,Zavala15,Lagos16b}, 
and so it is expected that low spin halos preferentially lead to the formation of galaxies 
with low spins.

The bottom panel of Fig.~\ref{LambdaHalo} shows  
 the distribution of spin parameters of the $z=1$ halos that contain the progenitors 
of the slow rotators at $z=0$. This shows that the spins of the halos hosting the slow rotators 
that never had mergers were already low $7$~Gyr ago, preventing the galaxies from reaching significant 
$\lambda_{\rm R}$. Interestingly, we see that on average the halos hosting these galaxies 
decrease their $\lambda_{\rm DM}$ from $\approx 0.025$ to $0.02$ from $z=1$ to $z=0$, which may be the cause 
for the systematic spinning down displayed by the slow rotators that never had mergers 
(solid lines in Fig.~\ref{SRMergersProgenitors}).

There is an overall weak positive correlation between $\lambda_{\rm r_{50}}$ and $\lambda_{\rm DM}$ for central 
galaxies (Fig.~\ref{LambdaHaloLambdaR}). However, when studying the incidence of slow rotators, a stronger 
correlation with $\lambda_{\rm DM}$ emerges.
This is investigated in Fig.~\ref{LambdaHaloFSR} for central galaxies
in the Ref-L100N1504 simulation at $z=0$ in three bins of stellar mass. 
The average $F_{\rm SR}$ of all central galaxies 
is shown as the solid line, 
while the subsamples of galaxies that had mergers 
and those that did not have mergers are shown as dashed and dotted lines, respectively.
The stellar mass bins were chosen to have $>500$ galaxies in each of the three samples above.

{The dependence of $F_{\rm SR}$ on $\lambda_{\rm DM}$ is close to monotonic with
$F_{\rm SR}$ decreasing with increasing $\lambda_{\rm DM}$. 
We find that galaxies that have had mergers at $z<2$ have a higher 
$F_{\rm SR}$ compared to galaxies that have not had mergers at fixed $\lambda_{\rm DM}$. 
The correlation is similarly tight for the different samples; i.e. jackknife errors are 
of a similar magnitude regardless of the merger history of galaxies. The top and middle panels of 
Fig.~\ref{LambdaHaloFSR} shows that galaxies that have not had mergers and are hosted by halos of 
low $\lambda_{\rm DM}$, have a $F_{\rm SR}$ that is similar or higher than that of the overall galaxy population 
at that stellar mass. Our results show the importance of the halo spin in determining slow rotation in 
central galaxies.}

\section{Discussion and conclusions}\label{conclusions}

Recent observational results from IFS surveys have reached 
contradictory conclusions regarding the effect of environment on the frequency  
 of slow rotators. The early work from ATLAS$^{\rm 3D}$ \citep{Cappellari11} 
concluded that the fraction of slow rotators increases steeply with stellar mass and
towards denser environments \citep{Emsellem11}. 
However, recent surveys that sample much larger numbers of galaxies have 
concluded that there is only a very weak or no dependence on environment once stellar mass is controlled for 
\citep{Brough17,Veale17,Greene17}. Here we used the \eagle\ and \Hydrangea\ simulations to explore this question 
and shed light onto the formation mechanisms of slow rotators.

We took special care in constructing IFS-like cubes for all of our simulated galaxies 
to measure the relevant quantities, $\lambda_{\rm R}$ and $\epsilon$, in a way that 
is more directly comparable to observations. 
We classify galaxies in \eagle\ and \Hydrangea\ that have stellar masses $>5\times 10^{9}\,\rm M_{\odot}$ 
into slow and fast rotators, using several observational criteria. We compare with the observations 
of \citet{Emsellem11}, \citet{Brough17} and \citet{Veale17} and find that our simulations reproduce 
 the dependence of the fraction of slow rotators, $\rm F_{\rm SR}$, on stellar mass relatively well.
We showed that by applying {a small error to our measurements of $\lambda_{\rm R}$} we recover excellent 
agreement with the observations at $M_{\rm stars}\lesssim 10^{11.5}\,\rm M_{\odot}$ (Fig.~\ref{FSRMass}). {At higher masses, 
we find a low frequency of slow rotators,  
$\approx 26$\%, while observations point to a much higher fraction 
 $>50$\% (\citealt{Oliva-Altamirano17}; \citealt{Brough17}). This discrepancy is likely due to BCGs in \eagle\ and 
\Hydrangea\ being overly massive for their halo mass and have star formation rates that are higher than 
observations. Continuing star formation 
is very efficient at spinning up galaxies, resulting in BCGs being mostly fast rotators.}

We explored the effect of environment in two ways: by separating centrals and satellites, 
and by studying the effect of halo mass on the distribution of galaxies in the $\rm F_{\rm SR}$-stellar mass plane.
We find that satellite galaxies are $49\pm15$\% more likely to be slow rotators
than centrals at stellar masses in the range $10^{10.8}\,\rm M_{\odot}-10^{11.5}\,\rm M_{\odot}$. At lower masses we find 
little differences in the overall populations of satellites and centrals (Fig.~\ref{FSRMass2}). However, when focusing on the 
passive population, we find that centrals of masses $M_{\rm stars}\lesssim 10^{10.7}\,\rm M_{\odot}$ 
are twice as likely to be slow rotators than satellites are (Fig.~\ref{FSRMass2Obs}). We interpret this as centrals 
undergoing quenching and morphological transformation simultaneously, while satellites 
can quench due to the environment they live in without changing morphology.

We separate satellites and centrals by the halo mass they reside in, and find 
a significant trend with halo mass for centrals galaxies, where 
$\rm F_{\rm SR}$ increases with increasing halo mass at fixed stellar mass (top panel of Fig.~\ref{FSRMass4}).
Satellite galaxies on the other hand show no dependence on halo mass once stellar mass 
in controlled for (bottom panel Fig.~\ref{FSRMass4}). However, the subsample of passive satellites shows a significant 
trend with halo mass, with $\rm F_{\rm SR}$ increasing with decreasing halo mass at $M_{\rm stars}<10^{10.5}\,\rm M_{\odot}$ 
(Fig.~\ref{FSRMass4b}). We speculate that satellite galaxies in low-mass 
halos, $M_{\rm halo}\lesssim 10^{13}\,\rm M_{\odot}$, require morphological transformation 
 to be quenched, while this is not the case in massive halos, $M_{\rm halo}\gtrsim 10^{14}\,\rm M_{\odot}$. 
{\citet{Correa17} presented 
an analysis of the connection between the bulge-to-total stellar mass ratio of \eagle\ galaxies 
with their colours. The authors concluded that satellite galaxies in the red sequence are more 
morphologically diverse compared to centrals, consistent with satellites quenching without 
having to transform morphologically.}
Note that the latter may not hold for low-mass galaxies (here we are only analysing 
galaxies with stellar masses $>5\times 10^{9}\,\rm M_{\odot}$). 
These are predictions that should be testable with the full catalogues of MaNGA \citep{Bundy15} and SAMI \citep{Bryant15} 
in combination with high-quality group catalogues \citep{Yang07,Robotham11,Saulder16}.

We use the extended merger tree information of \eagle,
as described in \citet{Qu17} and \citet{Lagos17}, to study the formation history of 
simulated galaxies. We find that there is a strong correlation between 
slow rotation and the incidence of dry mergers. 
Most galaxies ($\approx 60$\%) that have had at least one dry major merger in the last $10$~Gyr reside in the 
slow rotation region of the $\lambda_{\rm r_{\rm 50}}$-$\epsilon_{\rm r_{\rm 50}}$ plane.
Less frequent, but nonetheless common among slow rotators, are dry minor mergers.
Wet major and minor mergers are however more common in fast rotators (see Fig.~\ref{SRMergers}).
We find that the region of $\lambda_{\rm r_{\rm 50}}\gtrsim 0.7$ and $\epsilon_{\rm r_{\rm 50}}\gtrsim 0.6$
is almost exclusively occupied by galaxies that have not had any mergers with mass ratios $\ge 0.1$. 
Separating centrals and satellites, we find that dry major mergers are twice more common 
than any other merger with mass ratio $\ge 0.1$ in the population of central slow rotators, while for satellites 
dry minor and major mergers are the dominant form of mergers (Fig~\ref{SRMergersHistory}). 

By studying individual merger events, we find that dry major and minor mergers tend to be 
 associated with a net spin down of galaxies, while wet 
mergers can spin up galaxies very efficiently (Fig.~\ref{LambdaRvsMergers}).
We find that mergers have a cumulative effect, and galaxies undergoing 
successive minor mergers are more likely to spin down 
and become slow rotators. For comparison, galaxies that had $\ge 3$ mergers 
have an incidence of slow rotators of $30$\%, while this fraction decreases to 
$10$\% in galaxies that had one merger (not shown here).
{We also found a secondary effect of the orbital angular momentum on the 
remnant $\lambda_{\rm r_{\rm 50}}$ in the case of major mergers, 
in a way that lower orbital angular momentum leads to a larger decrease in 
 $\lambda_{\rm r_{\rm 50}}$ (Fig.~\ref{LambdaRvsMergers2}).}
Surprisingly, $\approx 30$\% of the slow rotators in \eagle\ have not had mergers with 
mass ratios $\ge 0.1$. Those galaxies tend to have been born in halos of low spins (Fig.~\ref{LambdaHalo}) and we find that they 
currently reside in halos with median spin at least twice smaller than the rest of the slow 
rotators and the overall galaxy population. 


\eagle\ shows that although the formation paths of slow rotators can be varied, as previously pointed out 
by \citet{Naab14} using a small sample of $44$ simulated galaxies, there are preferred formation mechanisms. 
Those are dry major mergers in the case of central galaxies, 
dry minor and major mergers in the case of satellites, and 
being formed in halos of small spins in the case of slow rotators that have not had mergers.

One limitation we found is that the most massive galaxies in \eagle\ and \Hydrangea, $M_{\rm stars}>10^{11.8}\,\rm M_{\odot}$, 
are preferentially fast rotators, in contradiction with observations. 
This is connected to them being overly massive for their halo mass and star-forming \citep{Bahe17,Barnes17}. 
All these features are indicative of AGN feedback not being strong enough at the 
highest masses. In addition, \eagle\ lacks the population of very flat galaxies, $\epsilon_{\rm r_{\rm 50}}>0.75$. 
This is most likely due to the modelling of the ISM and cooling adopted in \eagle, 
as gas is forced to not cool down below $\approx 8,000$~K, which corresponds 
to a Jeans length of $\approx 1$~kpc, much larger than the scaleheights of 
disks in the local Universe \citep{Kregel02}. 
This issue could be solved by including the formation
of the cold ISM.
This, however, does not affect the capability of 
our simulations to study slow rotators.
Overall, our results show that simulations like \eagle\ and \Hydrangea\ are extremely 
powerful as their resolution allows us to look at their internal kinematics, at the same 
time as having large statistical samples to distinguish preferred formation scenarios. 

\section*{Acknowledgements}

The authors thank Eric Emsellem, Luca Cortese, Sarah Brough, Thorsten Naab and the theory and computing group at ICRAR for fruitful discussions. 
We also thank the anonymous referee for a constructive and helpful report.  
CL also thanks Rodrigo Tobar for the technical help. 
CL is funded by a Discovery Early Career Researcher Award (DE150100618).
CL also thanks the MERAC Foundation for a Postdoctoral Research Award and Cardiff University for their visitor program.
YB received funding from 
the European Union's Horizon 2020 research and innovation programme under the Marie Sklodowska-Curie grant agreement No 747645.
STK and DB acknowledge support from STFC through grant ST/L000768/1. 
JvdS is funded under Bland-Hawthorn's ARC Laureate Fellowship (FL140100278).
This work used the DiRAC Data Centric system at Durham University, operated by the Institute for Computational Cosmology on behalf of the STFC DiRAC HPC Facility ({\tt www.dirac.ac.uk}). 
This equipment was funded by BIS National E-infrastructure capital grant ST/K00042X/1, STFC capital grant ST/H008519/1, and STFC DiRAC Operations grant ST/K003267/1 and Durham University. DiRAC is part of the National E-Infrastructure.
Support was also received via the Interuniversity Attraction Poles Programme initiated
by the Belgian Science Policy Office ([AP P7/08 CHARM]), the
National Science Foundation under Grant No. NSF PHY11-25915,
and the UK Science and Technology Facilities Council (grant numbers ST/F001166/1 and ST/I000976/1) via rolling and
consolidating grants awarded to the ICC.
We acknowledge the Virgo Consortium for making
their simulation data available. The \eagle\ simulations were performed using the DiRAC-2 facility at
Durham, managed by the ICC, and the PRACE facility Curie based in France at TGCC, CEA, Bruyeres-le-Chatel.
This research was supported in part by the National Science Foundation under Grant No. NSF PHY11-25915.
The Hydrangea simulations were in part performed on the German federal maximum performance computer HazelHen at 
the maximum performance computing centre Stuttgart (HLRS), under project GCS-HYDA / ID 44067 financed through the large-scale project Hydrangea 
of the Gauss Center for Supercomputing. Further simulations were performed at the Max Planck Computing and Data Facility in Garching, Germany.
Parts of this research were conducted by the Australian Research Council Centre of Excellence for All-sky 
Astrophysics (CAASTRO), through project number CE110001020.
This work was also supported by the Netherlands Organisation for Scientific
Research (NWO), through VICI grant 639.043.409.

\bibliographystyle{mn2e_trunc8}
\bibliography{KinematicQuenching}

\label{lastpage}
\appendix
\input{app_convergence}

\end{document}

%% file: app_convergence.tex
\section[]{Convergence tests}\label{ConvTests}

\subsection{Resolution convergence}\label{rescovapp}

We present convergence tests for the ellipticity and $\lambda_{\rm R}$ measurements
performed on galaxies with $M_{\rm stars}\ge 5\times 10^9\rm \, M_{\odot}$ at $r_{\rm 50}$. 
The stellar mass limit above was motivated by \citet{Lagos16b} as the stellar mass above which 
the stellar specific angular momentum of galaxies measured at $r\ge r_{\rm 50}$ converges. 
For our convergence test we use the run referred to as Recal-L025N0752 in S15, which 
corresponds to a volume of length $L=25$~cMpc and with $2\times 752^3$ particles, and that adopts the same sub-grid 
physics as the reference simulation used in this work (Ref-L100N1504 and Ref-L050N752; see Table~\ref{TableSimus}), 
but has parameters adjusted to fit the stellar mass function at $z=0$.
This is referred to as `weak convergence' test in S15. 
To allow for a fair comparison, we use the Ref-L025N0376, which has the same resolution, 
subgrid physics and parameters as the simulations in Table~\ref{TableSimus}, but with 
a box of length $L=25$~cMpc. 

\begin{figure}
\begin{center}
\includegraphics[trim=1mm 6mm 0mm 0mm, clip,width=0.43\textwidth]{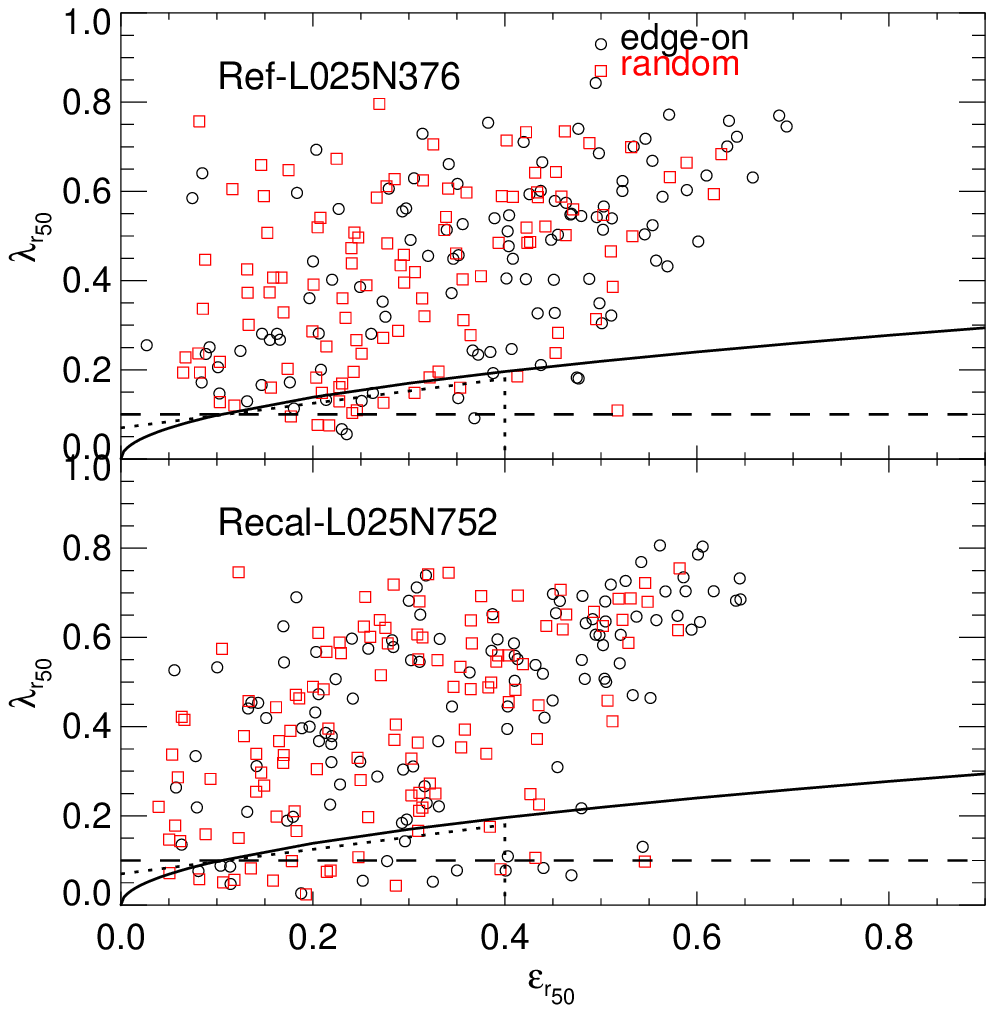}
\caption{$\lambda_{\rm r50}$ as a function of ellipticity for galaxies in the 
Ref-L025N0376 (top panel) and Recal-L025N0752 (bottom panel) simulations 
that have $M_{\rm stars}\ge 5\times 10^9\rm \, M_{\odot}$ at $z=0$. 
Circles and squares show galaxies seen edge-on and randomly, respectively.
The three lines correspond to different classifications of slow rotations, and are as in Fig.~\ref{LambdaREllip}.}
\label{ConvergenceTestEagle1}
\end{center}
\end{figure}

\begin{figure}
\begin{center}
\includegraphics[trim=1mm 5mm 0mm 0mm, clip,width=0.46\textwidth]{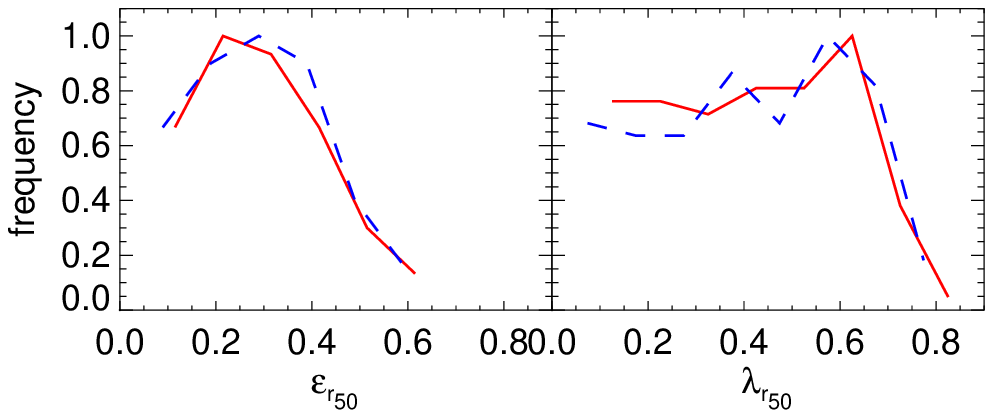}
\includegraphics[trim=1mm 6mm 0mm 0mm, clip,width=0.46\textwidth]{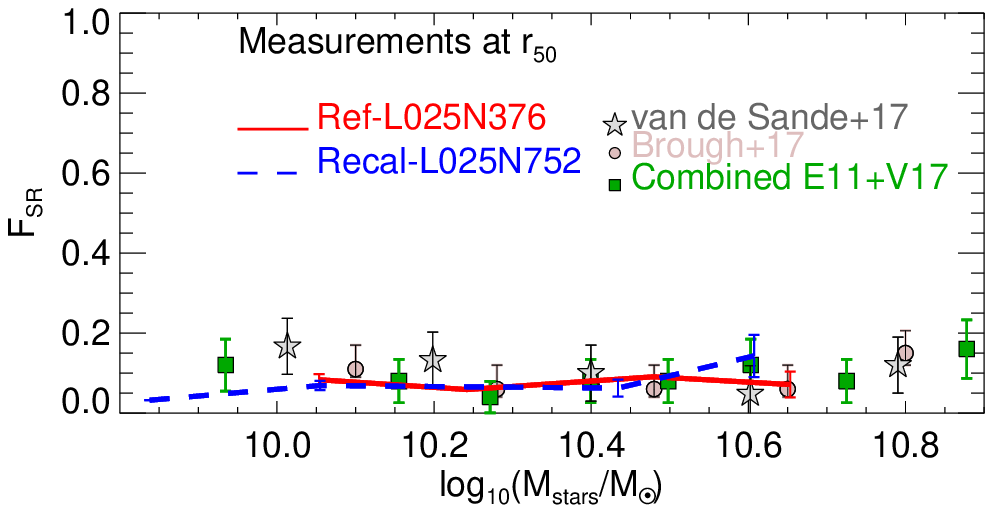}
\caption{{\it Top panel:} Distribution of $\lambda_{\rm r50}$ and $\epsilon$ 
for the galaxies in Fig.~\ref{ConvergenceTestEagle1}, {adopting random orientations}. Solid and dashed lines 
correspond to the Ref-L025N0376 and Recal-L025N0752 simulations, respectively. 
{\it Bottom panel:} $F_{\rm SR}$ using the \citet{Cappellari16} criterion 
as a function of stellar mass in the Ref-L025N0376 (solid line) and Recal-L025N0752 (dashed line) simulations. 
Error bars correspond to $1\sigma$ calculated
with $10$ jackknife resamplings in individual mass bins.}
\label{ConvergenceTestEagle2}
\end{center}
\end{figure}

Fig.~\ref{ConvergenceTestEagle1} shows $\lambda_{\rm r50}$ as a function of ellipticity for galaxies in the
Ref-L025N0376 and Recal-L025N0752 simulations. Both simulations occupy a similar parameter space, though with the 
Recal-L025N0752 simulation populating a bit more the high $\lambda_{\rm r50}$ area. 
This is better seen in the top panel of Fig.~\ref{ConvergenceTestEagle2}, which shows the 
distribution of $\lambda_{\rm r50}$ and ellipticity, measured adopting random orientations, for galaxies with 
$M_{\rm stars}\ge 5\times 10^9\rm \, M_{\odot}$ in both simulations. The simulations produce 
$\epsilon$ and $\lambda_{\rm r50}$ that are similar, with a slight tendency of the 
Recal-L025N0752 simulation {to produce galaxies that are more elongated}. 
Despite these differences, the fraction of slow rotators (bottom panel of Fig.~\ref{ConvergenceTestEagle2}) 
agrees very well, within the error bars. Since in this paper we are mainly concerned about the latter, 
we conclude that there is good convergence of the results presented throughout 
this manuscript. 

\subsection{Reference vs. AGNdT9 model}\label{modelcomp}

\begin{figure}
\begin{center}
\includegraphics[trim=1mm 5mm 0mm 0mm, clip,width=0.46\textwidth]{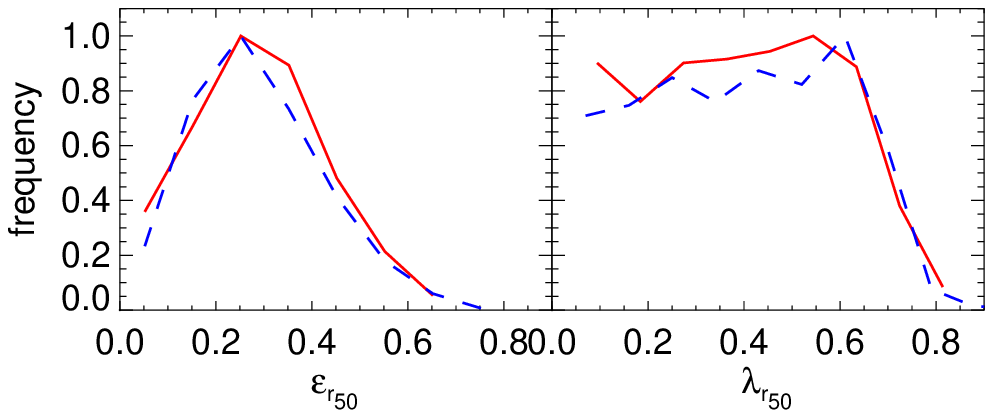}
\includegraphics[trim=1mm 6mm 0mm 0mm, clip,width=0.46\textwidth]{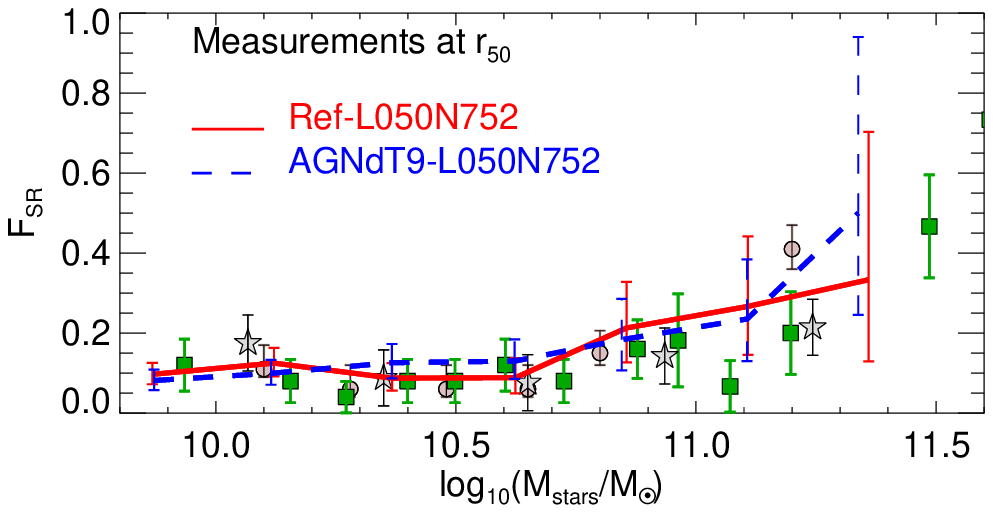}
\caption{As in Fig.~\ref{ConvergenceTestEagle2} but for the 
Ref-L050N0752 and the AGNdT9-L050N0752 simulations.}
\label{ConvergenceTestEagle3}
\end{center}
\end{figure}

The model adopted in \Hydrangea\ is the same as in the Reference \eagle\ runs, except 
for the temperature to which gas particles are heated by AGN. The reference \eagle\ model 
adopts $\Delta\,T_{\rm AGN}=10^{8.5}$~K and $C_{\rm visc}=2\pi$, 
while \Hydrangea\ adopts $\Delta\,T_{\rm AGN}=10^{9}\rm K$ and $C_{\rm visc}=2\pi\times 10^2$, 
with the purpose of decreasing the gas fraction 
is large groups and clusters. As part of \eagle, this model was run in the $50\,\rm (cMpc)^3$ 
box, and so here we compare these two models, fixing the box size, number of particles and initial 
conditions. We refer to these models as Ref-L050N0752 and the AGNdT9-L050N0752.

Fig.~\ref{ConvergenceTestEagle3} shows a comparison of $\epsilon_{\rm r_{50}}$ and $\lambda_{\rm r_{50}}$ 
in the Ref-L050N0752 and the AGNdT9-L050N0752 (top panel) and the fraction of slow rotators 
as a function of stellar mass at $z=0$ (bottom panel).
Both simulations show a similar $\epsilon_{\rm r_{50}}$ and $\lambda_{\rm r_{50}}$ distributions, and 
produce a similar $F_{\rm SR}-M_{\rm stellar}$ relation within the errorbars (bottom panel of Fig.~\ref{ConvergenceTestEagle3}).

\subsection{Convergence of kinematic measurements}\label{convbin}

\begin{figure}
\begin{center}
\includegraphics[trim=1mm 3.5mm 0mm 9mm, clip,width=0.45\textwidth]{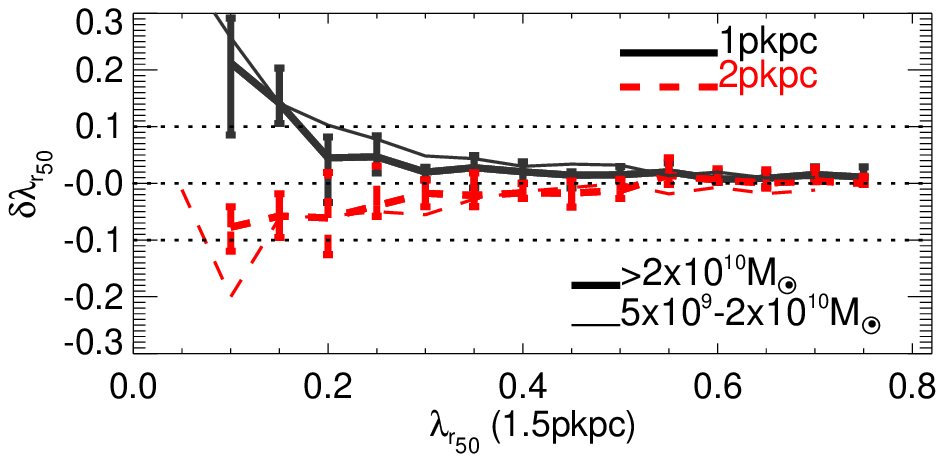}
\caption{The fractional variation of $\lambda_{\rm r_{50}}$ 
as a function of $\lambda_{\rm r_{50}}$ using bins of $1$~and~$2$~pkpc, as labelled (see Eq.~\ref{fracvar} for 
a definition of the fractional variation). Lines and error bars show the median and $25^{\rm th}-75^{\rm th}$ 
percentile ranges. {For clarity errorbars are shown only for one mass bin.}
In this test we use the Ref-L050N0752 simulation 
and show galaxies at $z=0$ {in two stellar mass bins, $5\times 10^9\,\rm M_{\odot}<M_{\rm stars}<2\times 10^{10}\,\rm M_{\odot}$ (thin lines) 
and $M_{\rm stars}\ge 2\times 10^{10}\,\rm M_{\odot}$ (thick lines)}.}
\label{ConvergenceTestEagle4}
\end{center}
\end{figure}

To create the r-band luminosity-weighted mock IFU-like cubes for each simulated galaxy we bin the 2D projected r-band luminosity map
as described in $\S$~\ref{kinematicmeasurements}. The chosen bin for the analysis of this paper 
is $1.5$~pkpc. Here we analyse the impact of this bin on 
$\lambda_{\rm r_{50}}$.
For this analysis,  
we use the Ref-L050N0752 simulation and select all galaxies that $z=0$ have 
$M_{\rm stars}\ge 5 \times 10^{9}\,\rm M_{\odot}$. We recompute all kinematic quantities 
but this time using bins of $1$~and~$2$~pkpc. We define the fractional variation $\delta$ 
of a kinematic quantity x in terms of the value obtained when we adopt a bin of $1.5$~pkpc,

\begin{equation}
	\delta x(\rm bin)= \frac{\left[x(\rm bin)-x(\rm 1.5~\rm pkpc)\right]}{x(\rm 1.5~\rm pkpc)}.
	\label{fracvar}
\end{equation}

In Fig.~\ref{ConvergenceTestEagle4} we show the impact the adopted bin has on the measurements of 
 $\lambda_{\rm r_{50}}$ in two bins of stellar mass. Adopting 
a smaller bin, closer to the softening length (see Table~\ref{TableSimus}),  
drives significant deviations in $\lambda_{\rm r_{50}}$ at low values of 
$\lambda_{\rm r_{50}}$, while adopting a larger bin does not have much of an impact. 
This is expected as choosing smaller bins, too close to the resolution limit, imply a much smaller 
number of particles per bin, which can produce significant errors in the determination of 
the kinematic parameters. 
{We find that the two bins of stellar mass reach similar results.}
Our results
show that choosing a bin of $1$~pkpc is not appropriate for a simulation of the resolution of \eagle, 
while choosing $1.5$ or $2$~pkpc has little impact on the kinematic measurements, implying good convergence. 
$\lambda_{\rm r_{50}}$ is converged to better than $7$\% 
for our adopted bin.